\documentclass[a4paper,useAMS,usenatbib,onecolumn]{mn2e}
\usepackage{times}
\usepackage{deluxetable}
\usepackage{longtable}
\usepackage{multirow}

\usepackage{url}
\usepackage{epsfig}
\usepackage{amsmath}
\usepackage{natbib}
\usepackage{graphicx}
\usepackage{aas_macros}
\begin{document}

\title[\textsc{DART-Ray}]{\textsc{DART-Ray}: a 3D ray-tracing radiative transfer code for calculating the propagation of light in dusty galaxies}

\author[Natale et al.]{\parbox{\textwidth}{G. Natale$^{1}$, C. C. Popescu$^{1,2}$, R. J. Tuffs$^{2}$, D. Semionov$^{1}$}\vspace{0.4cm}\\\\
\parbox{\textwidth}{$^{1}$Jeremiah Horrocks Institute, University of Central Lancashire, Preston, PR1 2HE, UK\\
$^{2}$Max Planck Institute f\"{u}r Kernphysik, Saupfercheckweg 1, D-69117 Heidelberg, Germany\\}}
\maketitle
\label{firstpage}

\begin{abstract}
We present \textsc{DART-Ray}, a new ray-tracing 3D dust radiative transfer (RT) code designed specifically to calculate radiation field 
energy density (RFED) distributions within dusty
galaxy models with arbitrary geometries. In this paper we introduce the basic algorithm implemented in \textsc{DART-Ray} which is based on a 
pre-calculation of a lower limit
for the RFED distribution. This pre-calculation allows us to estimate the extent of regions around the radiation sources within which
 these sources
contribute significantly to the RFED. In this way, ray-tracing calculations can be restricted to take place only within these regions, thus  
substantially reducing the computational time compared to a complete ray-tracing RT calculation. Anisotropic scattering is included in the code 
and handled in a similar fashion. Furthermore, the 
code utilizes a Cartesian adaptive spatial grid and an iterative method has been implemented to optimize the angular 
densities of the rays originated from each emitting cell. In order to verify the accuracy of the RT calculations performed by \textsc{DART-Ray}, 
we present results of comparisons with solutions obtained using the \textsc{DUSTY} 1D RT code for a dust shell illuminated by a central point source
and existing 2D RT calculations of disc galaxies with diffusely distributed stellar emission and dust opacity. Finally, we 
show the application of the code on a spiral galaxy model with logarithmic spiral arms in order to measure the effect of the spiral 
pattern on the attenuation and RFED. 

\end{abstract}
\section{Introduction}

Interstellar dust is of primary importance in determining the spectral energy distribution (SED) of the radiation 
escaping from galaxies at wavelengths ranging from the ultraviolet (UV) to the submillimetre (submm) and radio. Dust attenuates and 
redistributes the light, 
originating 
mainly from stars and, if present, from an active galactic nucleus (AGN), by either absorbing or scattering photons. The absorbed luminosity is then re-emitted 
in the infrared regime. The dust may be situated in complex geometries with respect to these sources, affecting the observed structure 
of the galaxy at each wavelength as well as its integrated SED.  
Modeling the propagation of light within real galaxies is thus a challenging task. Nevertheless, it is essential to do such modeling, 
if physical quantities of interest, 
such as the distribution and properties of the stellar populations and the interstellar medium (ISM) as traced by dust and the interstellar radiation fields, 
are to be derived from multiwavelength images and SEDs. \\ 

Taking advantage of the approximate cylindrical symmetry of galaxies, 2D dust radiative transfer (RT) models, such as the one presented by 
Popescu et al. (2011), already contain the main ingredients needed to predict integrated galaxy SEDs, average profiles, dust emission and 
attenuation for the case of normal star-forming disc galaxies. However, there are a number of reasons why 3D dust RT codes are desirable. First, spiral 
galaxies, although well modeled with 2D codes, show the presence of multiple and irregular features such as spiral structures, bars, 
warps and local clumpiness of the ISM. Also, galaxies may host a central AGN whose polar axis may not be aligned with that of the galaxy. 
For mergers or post-merger galaxies there is clearly no fundamental symmetry of the distribution of stars and dust. 
 Finally, solutions for the distribution of stars and ISM provided by numerical simulations of forming and evolving galaxies 
 generally require processing with a 3D RT code in order to predict the appearance in different bands. \\

The main challenge in realizing 3D solutions of the dust RT problem is the computational expense. The 
stationary 3D dust RT equation is a non-local non-linear equation: non-local in space (photons propagate within the entire 
domain), direction (due to scattering, absorption/re-emission) and wavelength (absorption/re-emission). Even using a relatively coarse 
resolution in each of the six fundamental variables, namely the three spatial coordinates, the two angles specifying the radiation 
direction and the wavelength, solving the 3D dust RT 
problem require an impressive amount of both memory and computational speed, at the limits of the capabilities of current computers. \\

Possibly the quickest way to calculate an image of a galaxy in direct and scattered light 
in a particular direction is by using Monte Carlo (MC) methods (including modern acceleration techniques, 
see Steinacker et al. 2013). 
There is a rich history of applications of MC codes to dust RT problems, 
starting with the pioneering works of e.g. Mattila (1970), Roark, Roark \& Collins (1974), Witt \& Stephens (1974) and Witt (1977). In the 
following decades the MC RT technique was further developed by many authors such as e.g. Witt, Thronson \& Capuano (1992), 
Fischer, Henning \& Yorke (1994), Bianchi, Ferrara \& Giovanardi (1996), Witt \& Gordon (1996) and Dullemond \& Turolla (2000). 
Nowadays, this method  
can be considered as the mainstream approach to 3D dust RT calculations (see e.g. Gordon et al. 2001, Ercolano et al. 2005,  Jonsson 2006,  
Bianchi 2008, Chakrabarti \& Whitney 2009, Baes et al. 2011, Robitaille 2011, but also see table 1 of Steinacker et al. 2013 
for a recent list of published 3D dust RT codes). 
The MC approach to dust RT consists of a simulation of the 
propagation of photons within a discretized spatial domain, based on a probabilistic determination of the location of emission 
of the photons, their initial propagation direction, the position where an interaction event (absorption or scattering) occurs and the 
new propagation direction after a scattering event. Thus, the MC technique mimics closely the actual processes occurring in nature
which shape the appearance of galaxies in UV/optical light. However, since it is based on a probabilistic approach to determine 
the photon propagation directions, an RT MC calculation does not necessarily determine the radiation field energy density (RFED) accurately in 
the entire volume of the calculation. 
The reason is that regions which have a low probability of being illuminated are reached by only few photons unless the total number of photons in 
the RT run is substantially increased. Nonetheless, in the case of disc galaxies, accurate calculation of radiation field intensities throughout the 
entire volume is needed, in particular for the calculation of dust emission. Indeed, far-infrared/submm observations of spiral galaxies show that most of the dust emission 
luminosity is emitted longwards of 100$\mu$m (see e.g. Sodroski et al. 1997, Odenwald et al. 1998, Popescu et al. 2002, 
Popescu \& Tuffs 2002, Dale et al. 2007, 2012, Bendo et al. 2012) through grains 
situated in the diffuse ISM which are generally located at very considerable distances from the stars 
heating the dust.

Another method to solve the RT problem in galaxies, alternative to the mainstream MC approach, is by using a ray-tracing 
algorithm. This method consists in the calculation of 
the variation of the radiation specific intensity along a finite set of directions, usually referred to as "rays". Ray-tracing algorithms
can be specifically designed to calculate radiation field intensities throughout the entire volume considered in 
the RT calculation. Also, it should be pointed out that MC codes already make large use of ray-tracing operations (see Steinacker et al. 
2013). It is thus interesting to pursue 
in the developing of pure ray-tracing 3D RT codes, which can be sufficiently efficient for the modelling of galaxies with 
3D arbitrary geometries, if appropriate acceleration techniques are implemented. Similar to MC codes, ray-tracing dust RT codes have 
had a rich history in astrophysics (see e.g. Hummer \& Rybicki 1971, Rowan-Robinson
1980, Efstathiou \& Rowan-Robinson 1990, Siebenmorgen et al. 1992, 
Semionov \& Vansevi\u{c}ius 2005, 2006). Application to analysis of 
galaxies started with the 2D code of Kylafis \& Bahcall (1987). Although originally implemented only for the calculation of optical 
images (see also Xilouris et al. 1997, 1998,1999), this algorithm was later adapted by Popescu et al. (2000) for the
calculation of radiation fields and was coupled with a dust emission model (including stochastic heating of grains) to predict the full
mid-infrared (MIR)/FIR/submm SED of spiral galaxies (see also Misiriotis et al. 2001,  Popescu et al. 2011). Thus far, extensions of the ray-tracing 
technique to 3D have been implemented but are specifically designed 
for solving the RT problem for star forming clouds (e.g. Steinacker et al. 2003, Kuiper et al. 2010), heated by few dominant  
discrete sources, rather than for very extended distributions of emission and dust as encountered in galaxies. \\

In this paper, we present \textsc{DART-Ray}\footnote{The name of the code can be seen as the acronym for ``Dust Adaptive Radiative Transfer Ray-tracing''
}, a new ray-tracing 
algorithm which is optimized for the solution of the 3D dust RT problem for 
galaxies with arbitrary geometries and moderate optical depth at optical/UV wavelengths\footnote{The actual version of the code 
does not consider the dust 
absorption/scattering of light emitted by dust at other positions (so called ``dust self-heating''). This effect can be neglected in 
case the galaxies are optically thin at infrared wavelengths}. The main challenge faced by this model is the construction of an efficient
algorithm for the placing of rays 
throughout the volume of the galaxy. In fact, a complete ray-tracing calculation between all the cells, used to discretise a model, 
is not a viable option, since it is by far too computationally expensive even for relatively coarse spatial resolution. 
Our algorithm circumvents the problem by performing an appropriate pre-calculation, whose goal is to provide a lower limit to the 
RFED distribution throughout the model. In this way, the ray angular density needed in the actual RT calculation can be dynamically 
adjusted such that the ray contributions to the local RFED are calculated only within the 
fraction of the volume where these contributions are not going to be negligible. \\

Furthermore, the code we developed can be coupled with any dust emission model. Applications of the 3D code for calculation of infrared
emission from stochastically heated dust grains of various sizes and composition, including heating of Polycyclic Aromatic Hydrocarbons 
molecules, utilizes the dust emission model from Popescu et al. (2011), and will be given in a future paper.  

The paper is structured as follows. In \S2 we provide some background information and motivation behind our particular ray-tracing 
solution strategy. In \S3 we give a technical description of our code. In \S4 we provide some notes on implementation and 
performance of the code. In \S5 we compare solutions provided by our code with 
those calculated by the 1D code \textsc{DUSTY} and the 2D RT calculations performed by Popescu et al. (2011). In \S6 we show the application 
of the code on a galaxy model including logarithmic spiral arms. A summary closes the paper. A list of definitions for the 
terms and expressions used throughout the paper can be found in Table \ref{tab_terms}.

\begin{table}
\begin{center}
\caption{Tables of terms and definition. The subscript $\lambda$ denotes a dependence from the wavelength of the radiation.}
\begin{tabular}{|p{4cm}|p{10cm}|}
\hline
Term &  Definition   \\ \hline
$A_{\rm{EM}}$ & Projected area of an emitting cell \\
$A_{\rm{INT}}$ & Projected area of an intersected cell \\ 
$f_L$ & Input parameter needed to set the threshold value for $\Delta L_\lambda$ below which scattering iterations are stopped \\
$f_U$ & Input parameter needed to set the threshold value for $\delta U_\lambda$ below which rays are stopped \\
$g_\lambda$ & Henyey-Greenstein scattering phase function parameter \\ 
$I_\lambda$ & Radiation specific intensity  \\
$<I_\lambda>$ & Average of $I_\lambda$ over the path crossed by a ray within an intersected cell \\ 
$I_{\lambda,i}$ & Radiation specific intensity of a ray before crossing a cell $i$ \\ 
$I_{\lambda,\rm{esc}}$ & Radiation specific intensity of a ray once it has reached the model border \\
$I_{\lambda,\rm{ext}} $ & Amount of radiation specific intensity of a ray extincted within an intersected cell \\ 
$\delta I_{\lambda,\rm{sca}}(\theta,\phi)$ & Ray contribution to the specific intensity of the radiation scattered into direction 
($\theta,\phi$), see Fig.\ref{i_sca}  \\
$j_\lambda$ & Volume emissivity, that is, the luminosity emitted at a certain position per unit volume, unit solid angle and unit wavelength
interval (also $j_{\lambda,c}$ when refereed to the average volume emissivity within a cell)  \\
$\kappa_\lambda$ & Dust extinction coefficient per unit dust mass (also $\kappa_{\lambda,\rm{abs}}$ or $\kappa_{\lambda,\rm{sca}}$ when referred 
to the extinction coefficients due to dust absorption or scattering) \\
$L_\lambda$ & Total stellar luminosity density of the model \\
$\Delta L_\lambda$ & Amount of luminosity still to be processed during scattering iterations \\ 
$L_{\lambda,\rm{ray}}$ & Luminosity associated with a ray beam \\ 
$N_{\rm{rays}}$ & Input parameter specifying the minimum number of rays crossing a cell within the fully sampled region \\
$\Phi_\lambda$ & Scattering phase function  \\
$\rho$ & Dust mass density (also $\rho_c$ when referred to the average dust mass density within a cell)  \\
$\Delta r$ & Ray path within an intersected cell \\ 
SCATT\_EN($\theta,\phi$) & Array storing the luminosity of the radiation scattered by dusty cells into a finite set of solid angles \\   
$\tau_\lambda$ & Optical depth  \\
$U_\lambda$  & Radiation field energy density (also referred to as RFED) \\
$\delta U_\lambda$ & Contribution to the local RFED carried by a ray (also $U_{\lambda,\rm{INT}}$ when referred to the 
contribution to an intersected cell RFED)\\   
$U_{\lambda,\rm{LL}}$  & Lower limit to the RFED \\
$U_{\rm{TEMP}}$ & Temporary array used to store RFED contributions from an emitting cell throughout the model \\
$U_{\lambda, \rm{FINAL}}$ & Array storing the RFED distribution which is output by the code \\  
$V_{\rm{INT}}$ & Intersected cell volume \\ 
$\omega_\lambda$ & Albedo \\
$\Omega_{\rm{HP,EM}}$ & Solid angle associated with the HEALPix spherical pixels used to define the directions of
rays from an emitting cell \\  
$\Omega_{\rm{HP,INT}}$ & Solid angle associated with the HEALPix spherical pixels used to define the directions of the radiation scattered 
by an intersected cell \\ 
$\Omega_{\rm{HP,MS}}$ & Solid angle associated with an HEALPix main sector (see Fig.\ref{hpix_sphere}) \\ 
$\Omega_{\rm{INT}}$ & Solid angle subtended by projected area of the intersected cell $A_{\rm{INT}}$ (see Fig.\ref{i_sca} \\ 
``Dusty cell'' & A cell where the average value of dust density is higher than zero \\
``Dust self-heating'' & The dust absorption of radiation emitted by dust at other positions (not included in this version of the code) \\
``Emitting cell'' & A cell where the average value of the stellar light or scattered light volume emissivity is higher than zero \\
``Escaping radiation'' & The direct or scattered stellar radiation propagating outside the borders of the volume considered 
in the RT calculation \\
``Full sampling'' (of a region)  & The process of launching enough rays from a source, such that all the cells within a region are 
intersected by multiple rays \\
``Intersected cell'' & A cell intersected by a ray \\ 
``Leaf cell'' & A cell of the 3D adaptive grid which is not further subdivided \\
``Lost luminosity'' & Amount of stellar luminosity not considered in the RT calculation (to be kept low to guarantee approximate energy balance,
see \S4) \\ 
MC & Monte Carlo \\ 
RFED  & Radiation Field Energy Density \\
RT & Radiative Transfer \\
``Source influence volume'' & The fraction of model volume within which a source contributes significantly to the RFED \\
``Volume emissivity'' & see $j_{\lambda}$ \\ 

\hline

\end{tabular}
\label{tab_terms}
\end{center}
\end{table}

\section{Background and Motivations}
In this section we describe the basic characteristics of the time-independent 3D dust RT equation, briefly introduce the ray-tracing 
approach  used in our code and provide the main motivations behind our solution strategy. Finally, we describe the main steps of the new 
 algorithm we present in this work. A much more technical description of our code can be found in \S3. 

\subsection{Dust continuum 3D Radiative Transfer: Ray-tracing and Solution Strategy}

Given an input distribution of stellar luminosity and dust mass, solving the RT problem requires in principle 
the resolution of the following equation for the specific intensity $I(\lambda, \mathbf{x,n})$, which represents the luminosity 
per unit area, solid angle and wavelength interval propagating at point $\mathbf{x}$ into the direction $\mathbf{n}$:
\begin{equation}
 \mathbf{n}\nabla_{\mathbf x} I_\lambda(\mathbf{x,n}) = -k_\lambda(\mathbf{x})\rho(\mathbf{x})\left[I_\lambda(\mathbf{x,n}) \\
 - \omega_\lambda\int_\Omega{\Phi_\lambda(\mathbf{n,n'}) I_\lambda(\mathbf{x,n'})d\Omega'}\right]+j_\lambda(\mathbf{x}) 
\label{rf_equ}
\end{equation}
where $k_\lambda(\mathbf{x})$ is the total extinction coefficient per unit mass of dust (including both absorption and 
scattering), $\rho(\mathbf{x})$ is the dust mass density, $\omega_\lambda$ is the albedo, defined such that 
$\omega_\lambda\times k_\lambda(\mathbf{x})$ gives the fraction of extinction due to 
scattering, $\Phi_\lambda(\mathbf{n,n'})$ is the scattering phase function, which gives the probability for radiation coming from direction 
$\mathbf{n'}$ to be 
scattered into direction $\mathbf{n}$, and $j_\lambda(\mathbf{x})$ is the distribution of stellar volume emissivity\footnote{
Throughout the 
text by ``volume emissivity''
we will always mean the luminosity per unit volume per unit solid angle and per unit wavelength interval of the stellar radiation 
at each position. To not be confused with the emissivity coefficient used to characterise the emission properties of e.g. gas or dust.}. 
The first term on the 
right-hand side 
of Eq.\ref{rf_equ} acts to reduce the radiation specific intensity $I(\lambda, \mathbf{x,n})$ by a 
quantity that is proportional to the radiation intensity itself. The second term instead acts as a source term and gives a positive contribution to the 
radiation intensity by adding the light coming from all directions to point $\mathbf{x}$ and then scattered into the direction 
$\mathbf{n}$. The third term gives a positive contribution to the propagating radiation specific intensity, which is due to stellar 
emission and it is assumed to be isotropic at each position $\mathbf{x}$. 
Since in 3D RT there are six independent variables, namely wavelength, three spatial coordinates and two angular directions, the solution 
vector for $I_\lambda(\mathbf{x,n})$ can be extremely large, making the problem very challenging also from the point of view of memory requirements, 
apart from the difficulty of solving the integro-differential equation itself in three dimensions.
Note also that we did not include a term that represents the re-emission by dust, important at infrared wavelengths. The resolution 
algorithm presented in this work is designed only to handle the propagation of direct and scattered light from stellar populations and 
does not consider the self-heating of dust.  \\    

Instead of seeking to obtain a solution for $I_\lambda(\mathbf{x,n})$, 
the main aim of this work is to construct an algorithm optimized to derive the radiation field energy 
density $U_\lambda$ (hereafter also refereed to as RFED) at each position 
$\mathbf{x}$, in order to be able to calculate successively the dust emission spectra assuming local energy balance between dust 
radiative heating and emission\footnote{For example, the latter condition for a single grain stochastically heated can be 
expressed as:
$$
\int Q_{\lambda,abs} \int B_\lambda(T)P(T)dT d\lambda = (c/4\pi) \int Q_{\lambda, abs} U_\lambda d\lambda
$$
where $Q_{\lambda,abs}$ is the grain absorption efficiency at wavelength $\lambda$, $B_{\lambda}(T)$ is the Planck function calculated for dust temperature $T$ and 
$P(T)$ is the probability for the dust grain to have a temperature equal to $T$.  Numerical methods, such as the one presented in 
Guhathakurta \& Draine (1989), allow us to derive $P(T)$ and therefore the dust emission spectra, once the absorption efficiency 
 $Q_{\lambda,abs}$ and the RFED $U_{\lambda}$ are known..} 

In terms of $I_\lambda(\mathbf{x,n})$, one can express $U_{\lambda}$ as:

\begin{equation}
 U_\lambda(\mathbf{x})=\frac{\int{I_\lambda(\mathbf{x,n})d\Omega}}{c}
\label{en_equ}
\end{equation} 

In order to calculate $U_\lambda(\mathbf{x})$ at a specific point $\mathbf{x}$, one can simply sum up the contributions 
$\delta U_\lambda$ 
provided by the radiation coming from all the emitting sources to the value of $U_\lambda(\mathbf{x})$ at that position. 
A numerical method to calculate $U_\lambda(\mathbf{x})$ at any position can be implemented by considering 
``rays'' originating from each emitting source and propagating throughout the whole volume considered in the calculation. 
Along each ray path one follows the variation of the radiation intensity and   
one can thus calculate the $\delta U_\lambda$ contributions at a finite set of positions (this solution 
technique is among those known as ``ray-tracing'' methods). More specifically, a solution algorithm one could use to derive the 
distribution of $U_\lambda(\mathbf{x})$ for a single wavelength $\lambda$, given an input 3D distribution of stellar luminosity and 
dust mass, is the following. 

First, the entire model is subdivided in an 
adaptive grid of cubic cells and to each cell one assigns the average values for both the dust density and 
stellar volume emissivity within the cell volume. A cell for which the average value of stellar volume emissivity is 
higher than zero can be treated approximately as a discrete radiation source. In the following, we will refer to this kind of cell as an 
``emitting cell''. Similarly, cells with average dust density higher than zero will be referred to as ``dusty cells''. 

Then, 
one performs ray-tracing for a large set of 
directions originating from the centres of the emitting cells. That is, one follows the variation of the specific intensity
$I_\lambda(\mathbf{x,n})$ from the cell centres along rays corresponding to each direction $\mathbf{n}$. 
While following a ray, one considers the increase of radiation intensity $I_\lambda(\mathbf{x,n})$ due to the stellar volume 
emissivity in the 
cell originating the ray but not in the intersected cells, where only the decrease of intensity due to dust absorption and scattering is 
considered. This allows us to calculate separately the contributions $\delta U_\lambda$ provided only by the emitting cell originating 
the ray to the final value of $U_\lambda(\mathbf{x})$ in all the cells intersected by the same ray\footnote{This approach can be 
redundant because the rays originating from different cells can often go through almost the same paths. On the other side, 
it is convenient because it easily allows the implementation of the acceleration techniques presented in this work.}. When a ray 
intersects a cell $i$ different from the original cell, the new value of the specific intensity will then be:
\begin{equation}
I_{\lambda,i+1}=I_{\lambda,i}e^{-k\rho_{c,i}\Delta r}
\label{i_new_equ}
\end{equation}
where $\rho_c$ is the cell dust density and $\Delta r$ is the \emph{crossing path}. For each ray-cell intersection one calculates 
an appropriate average of the value of $I_\lambda$ within the ray crossing path and, thus, the 
contribution $\delta U_\lambda$ to the local value of $U_\lambda(\mathbf{x})$ by using a discrete version of Eq. \ref{en_equ}.
When all the rays from one emitting cell have been processed, 
the ray-tracing is performed from another emitting cell and so on until all the emitting cells have been considered. 
Scattered radiation, whose intensity in a finite number of directions has also been stored locally for each ray-cell intersection, 
is then processed in a similar fashion. 

If one creates a grid sufficiently fine in resolution and launches a sufficiently large number of rays, 
the calculated cell RFED values are very close to the exact values of 
$U_\lambda(\mathbf{x})$ at the cell 
centres and can be used to calculate the dust heating. Furthermore, the described 
procedure is extremely flexible and capable 
to handle completely arbitrary 3D distributions of stellar luminosity and dust mass. Unfortunately, the implementation of the procedure in the simple form 
described above is too computationally expensive (scaling approximately as $N^{5/3}$, with $N$ being the total number of cells, in the case of 
uniform spatial sampling), 
considering also that the calculation has to be performed in an iterative 
way for the scattered light and 
for different wavelengths.  
Nonetheless, for well mixed emitters--absorbers 
geometries, such as in the case of galaxy dust--stellar distributions, the full ray-tracing calculation for rays propagating throughout
 the entire model from each emitting cell does not have to be necessarily performed to obtain a reasonably accurate solution 
 for $U_\lambda(\mathbf{x})$. 
 
\begin{figure}
\includegraphics[scale=0.8]{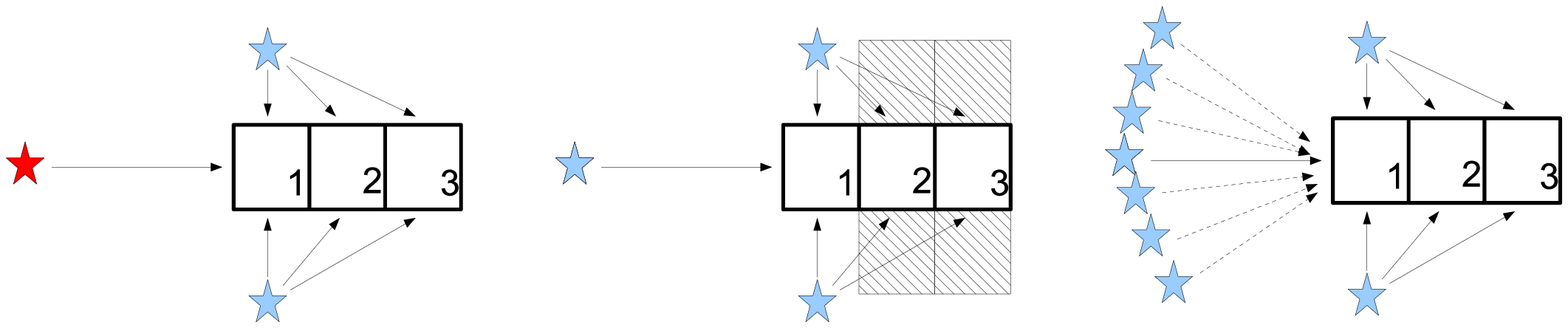}
 \caption{Examples of stars and dust geometries where the criteria to identify the limit of a source influence volume works (left panel)
 or can fail (middle and right panel). Left: The ray coming from the distant source does not contribute significantly to the RFED in cell 1, 
 which is illuminated mainly by other sources.  As a consequence, the same ray does not contribute significantly also to the cells 
 beyond cell 1; Middle: As in the left panel, the ray from the distant source
 does not contribute significantly to the RFED in cell 1. However, in this particular geometry, its contribution to the RFED in the 
 cells beyond cell 1 can be not negligible because the emission from the other sources is highly attenuated by dust (dashed cells); 
 Right: the distance source is part of a large emitter distribution, whose individual RFED contributions to cell 1 are very small but  
the cumulative contribution can be not negligible.} 
\label{en_crit}
\end{figure}

In fact, each radiation source in a general RT problem not necessarily contributes significantly to $U_\lambda$ at each 
position within the volume considered for the RT calculation but often only within a fraction of it, which we call 
the ``source influence volume''. In the ray-tracing method described above, if one knew in advance the extent of the influence 
volume for each emitting cell, 
it would be possible to reduce the number of calculations by simply performing ray-tracing only within this volume. 
Unfortunately, it is not possible to know this information a priori, since the final values of $U_\lambda$ 
 at each position are available only \emph{after} performing the RT calculation. However, one can estimate in a conservative way 
the extent of the influence volume for a given emitting cell. The basic idea is to identify a point at some distance from the 
emitting cell where it can be proved that the contribution $\delta U_\lambda$ carried by a ray is negligible compared to the local final value
of $U_\lambda$. If this is the case, this can imply that the emitting cell will 
not contribute significantly to $U_\lambda$ beyond that position, which is already outside the 
influence volume of the emitting cell [see Fig.\ref{en_crit} for simple examples where this criteria will work (left-hand panel) or 
can fail (middle and right-hand panels) if not properly applied]. 

\begin{figure}
\centering
\includegraphics[scale=0.8]{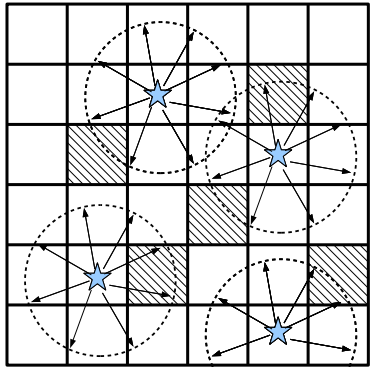}
\caption{First estimate of the RFED. The radial ray-tracing is performed only until a limit optical depth or distance. Dashed squares
denote cell containing dust.}
\label{fe_ray_prop}
\end{figure}

In practice, one can implement this method in the following way. First, one calculates a 
lower limit $U_{\lambda,\rm{LL}}$ 
for $U_\lambda$ within the entire volume by performing ray-tracing from each emitting cell only until a certain arbitrary distance 
(see Fig.\ref{fe_ray_prop}). Once the lower limit for 
$U_\lambda$ is 
calculated, one can start the RT calculation again from the beginning but this time is able to check if a 
contribution $\delta U_\lambda$,
carried by a ray to a particular cell, is going to be significant or not. In fact, if for a certain intersected cell $\delta U_\lambda$
is negligible compared to the local value of the lower limit $U_{\lambda,\rm{LL}}$, then it will be negligible also compared to the 
final value of $U_\lambda$ at that position. The actual implementation of this method consists in checking at each cell intersection if 
$\delta U_\lambda < f_U\times U_{\lambda,LL}$ with $f_U$ equal to a very small number to be chosen appropriately (see \S4 for discussion on 
this point). When this condition
is realized and if the chosen value of $f_U$ is sufficiently small, then 
the contributions $\delta U_\lambda$ will also be negligible for all the cells beyond that position in the same ray direction.

\begin{figure}
\centering
\includegraphics[scale=0.8]{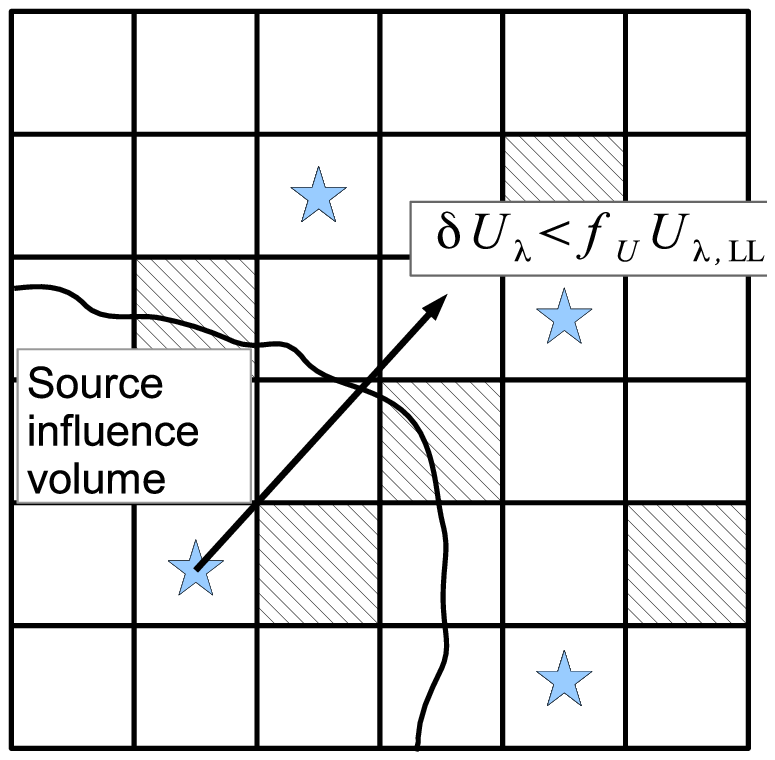}
\caption{Ray-tracing calculation from one source until $\delta U_\lambda < f_U \times U_{\lambda,LL}$. The position when this condition 
is realized should be outside the source influence volume. Thus, one can stop the ray-tracing calculation at that position.}
\label{fig_step1}
\end{figure}

This implies that, once $\delta U_\lambda < f_U\times U_{\lambda,LL}$ at a certain distance from an emitting cell 
along a ray-path,
one can stop the ray-tracing calculation for that particular ray at that position. In fact, in that case, the intersected cell is 
already outside the influence volume of the emitting cell originating the ray (see Fig.\ref{fig_step1}). In this way, the total 
amount of calculations to be performed can be substantially reduced, making it feasible to use a modified version of 
the above ray-tracing algorithm to infer the $U_\lambda$ distribution within
complex dust/stars structures as those observed in galaxies. In the following subsection, we provide a simplified description
of the RT algorithm we have developed based on this approach. \\

\subsection{Basic Description of the DART-Ray Algorithm}
In the following we provide a simple description of the three steps performed by the algorithm which implements the solution strategy 
outlined above to calculate the RFED distribution $U_\lambda$ at a single wavelength $\lambda$. This 
description assumes that a grid of cells subdividing the entire model has already been created. 
A complete technical description of the code will be given in Section \S3.   \\

\textit{First Step: Calculation of a Lower Limit for $U_\lambda$}\\
The first step consists of ray-tracing from each emitting cell adopting a ray angular density such that \textit{all} the cells 
within a certain radius or a 
certain optical depth from each emitting cell are intersected by multiple rays, as shown in Fig.\ref{fe_ray_prop} (hereafter, 
we will refer to the process of intersecting all cells in a certain region with multiple rays as ``fully sampling'' that region). 
The specific values for the limit radius and optical depth can be specified in the input. 
They should be large enough to let the rays cover a relatively large fraction of the entire model but small enough to avoid a too long computational time 
in this step. 
The inferred contributions $\delta U_\lambda$ to 
the value of $U_\lambda$ in each cell crossed by rays are accumulated into an array $U_{\lambda,\rm{LL}}$ (LL stays as ``lower limit''). 
As said before, the inferred RFED distribution $U_{\lambda,\rm{LL}}$ represents only a lower limit for the final value  $U_\lambda$ 
since there could be other emitting cells, at distances 
larger than those adopted limits, whose contribution to $U_\lambda$ at each particular point has not been considered yet. 
In addition, scattered light contributions have also been neglected at this point. \\

\textit{Second Step: Processing of Source Direct Light}\\
In the second step the ray-tracing procedure is repeated again from the beginning but this time the rays fully sample the regions 
around each emitting cell, until the ray contribution $\delta U_\lambda$ for an intersected cell become smaller than a very small 
fraction of the lower limit $U_{\lambda,\rm{LL}}$, 
that is until $\delta U_\lambda < f_U \times U_{\lambda,LL}$ with $f_U$ being a very small number (see \S4 for more details 
about how to choose an appropriate value for $f_U$). When this condition is realized, 
it means that the position reached by the ray is outside the emitting cell influence volume (see Fig.\ref{fig_step1}) and 
the final value of $U_\lambda$ in the crossed cell is contributed mainly by emitting cells different from the emitting cell 
originating the ray\footnote{
Note that, as shown in the right-hand panel of Fig.\ref{en_crit}, there could be emitting cells which are 
part of a large emitting cell distribution, whose individual RFED 
contributions $\delta U_\lambda$ might be lower than the 
assumed energy density threshold $f_UU_{\lambda,LL}$, but the cumulative contribution 
is not negligible. Overlooking the presence of this kind of cells can be avoided by choosing an appropriate 
value for the constant $f_U$. However, in cases such as that of an extended distribution of uniform volume emissivity within an 
optically thin system, the value of $f_U$ to be chosen, in order to reach an accurate solution, could be extremely low. In those cases, 
there might be no substantial advantage in terms of speed by using the presented algorithm, since the majority of emitting cells 
contribute significantly to the RFED at each position in the entire volume.}. 
 Therefore, to the purpose of calculating the RFED distribution $U_\lambda$, there is no reason to 
proceed with a full sampling of the cells beyond the limit determined in this way. 
Apart from storing the values of the RFED contributions $\delta U_\lambda$, after each ray crosses a cell, the scattered energy 
information are also stored for that cell. That is, the luminosity scattered within a discrete set of solid angles is stored for each 
cell containing dust. These values will be used in the next step. 
Before going to the third step, the value of $U_{\lambda,\rm{LL}}$ is updated with the current estimation for $U_\lambda$. \\

\textit{Third Step: Scattering Iterations}\\  
After the direct light has been processed in the second step, a third step is started where there is a series of iterations to process 
the scattered 
radiation. Each cell which originates scattered light is treated as an emitting cell exactly in the same way as in step 2.
Ray-tracing is performed from each dusty cell by fully sampling all the volume surrounding the cell until the contribution
$\delta U_\lambda$ is negligible compared to a small fraction of $U_{\lambda,\rm{LL}}$, the lower limit for $U_\lambda$ updated with the new
value of $U_\lambda$ found in step 2. Again, scattering information 
are stored as well after each ray crossing and this higher order scattered radiation intensity is processed in successive iterations. 
These scattering iterations continue until the vast majority of the luminosity of the system has been either absorbed by dust or has 
escaped outside the model. That is, until $\Delta L_{\lambda} < f_L L_{\lambda}$, where $L_{\lambda}$ is the total stellar luminosity 
emitted within the model, $\Delta L_{\lambda}$ is the amount of luminosity still to be processed (that is, not absorbed or escaped yet) 
and $f_L$ is a parameter to be set in the input. \\   

The disadvantage of this procedure is 
that many of the calculations performed in the first step are repeated once again in the second step. 
The advantage is that the number of calculations avoided can be very high, thus reducing significantly the total calculation
time for those geometries where the influence volumes of the emitting cells are only a small fraction of 
the total volume. 
Further characteristics of the code, 
not mentioned in the simplified algorithm above, include the adaptive and directional-dependent angular density of the rays (see \S3.2) and the methods 
implemented to calculate the specific intensity of the radiation escaping outside the model in a finite set of directions 
(see Section \S3.4). 

\section{The Ray-Tracing 3D Code: Technical Description}
In this section we provide an extensive description of the code we have developed. The code consists of two main programs performing 
the adaptive grid creation and the RT calculation respectively.  In the following subsections we describe the adaptive 
grid creation (\S3.1), the basic ray-tracing routine (\S3.2) and each of the three 
steps of the RT algorithm in detail (\S3.3). Finally, we describe the methods implemented to derive the escaping radiation 
specific intensity (\S3.4). 

\subsection{Adaptive Grid Creation}
Given an input spatial distribution of dust mass and stellar luminosity (either defined by analytical formulae or in a tabulated form), 
an adaptive grid is created in a way such that the spatial resolution is higher in regions where the radiation field intensity is 
expected to vary in a more rapid way, such as those where the density of dust is higher.  
A parent cell of cubic shape, enclosing the entire model, is subdivided 
in 3x3x3=27 child cubic cells of equal size\footnote{We use a grid refinement factor equal to 3, since it has the advantage of 
preserving the position of cell centres
 after each cell subdivisions. For technical reasons, this facilitates the inclusion in the grid of central emitting point sources 
 such those in the solutions described in the next section.}. Then, the average dust density and stellar volume 
 emissivity are calculated within the volume of  
each newly created cell, together with an estimation of their variation within each cell.  
Further cell subdivision proceeds for those child cells which do not satisfy user-defined criteria 
and those cells become parents of even smaller child cells. After that, the estimation of the cell dust density/stellar 
volume emissivity and the 
cell subdivision 
procedure are performed for the new cells and so on. In this way, a tree of cells is constructed iteratively until the chosen criteria 
are satisfied for all 
the smallest cells within each original parent cell. Also, in order to obtain a smooth variation of the grid resolution, further cell 
subdivision is performed 
to avoid differences in cell subdivision level higher than one between neighbour cells.  
The cells which have not been further subdivided, after the entire grid creation has been completed, are called the leaf cells. 

The criteria for cell subdivision 
should be chosen in order to obtain both numerical accuracy and an adequate coverage of the RFED distribution within the model. 
In order to achieve a good 
numerical accuracy, it is important to have leaf cells with 
small total optical depths and with small gradients of dust density and stellar volume emissivity within the cells.
For example, typical conditions for a cell to be a leaf cell could be:
\begin{equation}
 \tau_\lambda << 1
\end{equation}

\begin{equation}
 \frac{\Delta\rho}{\rho_c} < 0.5 
 \label{delta_rho}
\end{equation}
where $\tau_\lambda=k_\lambda\rho_c l_c$ is the total cell optical depth, $\Delta \rho$ is an estimate of the dust density 
variation within a cell and
$\rho_c$ is the average cell dust density. An equivalent condition, as the one expressed by Eq. \ref{delta_rho}, has to be fulfilled by 
the cell stellar volume emissivity $j_{\lambda,c}$. These conditions allow us to use 
the simple expression given by Eq. \ref{i_new_equ} to calculate the variation of the specific intensity within a cell to a good 
degree of accuracy. 
However, note that an additional constraint is the maximum allowed number of subdivision levels NLVL\_MAX, which has to be chosen in the 
input as well. 
The value of this 
parameter should be such to guarantee that the input cell parameter requirements, such as those above, are valid for all or at least the 
vast majority of cells in the model and, 
at the same time, avoid to create models with too many cells. It is desirable to keep the number of cells as low as possible, since it 
is one of the main 
parameters affecting the total computational time. After creating the grid, the program creates a table where the user can read the 
maximum and average cell 
parameters and, thus, quickly verify to which degree the input cell requirements have been fulfilled. 

All the information which define the grid is printed on a file that can be read from the RT program. This includes:\\
-- cell id number\\
-- position of cell centre in Cartesian coordinates\\
-- cell child number : which is equal to ``-1'' if the cell is a leaf cell or to the id number of the first child cell created during 
the cell subdivision\\ 
-- cell index number : binary code expressing the position of the cell within the cell tree \\
-- cell size $l_c$ \\
-- cell dust density $\rho_c$\\
-- cell stellar volume emissivity $j_{\lambda,c}$\\ 


\subsection{The basic Ray-Tracing Routine}
\begin{figure}
\centering
\includegraphics[trim=0cm 0cm 0cm 1cm, scale=0.4, clip=true]{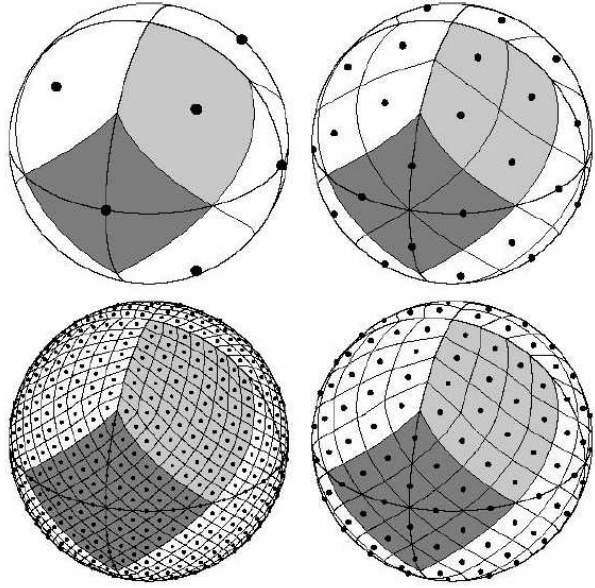}
\caption{Healpix sphere at different angular resolution. The spherical pixels in the upper left sphere are those mentioned 
as ``HEALPix main sectors'' in the text. Figure from Gorski et al. (2005) (reproduced by permission of the AAS).}
\label{hpix_sphere}
\end{figure}

In this subsection we describe the basic ray-tracing calculation for rays departing from an emitting cell and propagating 
throughout the model. 
This is the core routine used by the RT algorithm described in the next subsection. 
Given an emitting cell, rays are casted into 
multiple directions defined by using the HEALPix sphere pixelation scheme (Gorski et al. 2005, see also e.g. Abel \& Wandelt 2002 and 
Bisbas et al. 2012 for other applications in RT codes). In this scheme a sphere is divided in 
12 
main sectors of equal size, 
which can be further subdivided in smaller spherical pixels (see Fig.\ref{hpix_sphere}). By assuming that an HEALPix sphere is 
centred on the emitting cell 
centre,  the lines connecting the cell centre to the centre of the HEALPix spherical pixels define directions along which one can follow the variation of the specific 
intensity within the model. The use of the HEALPix routines PIX2ANG\_NEST (translating pixel index numbers into spherical coordinates) and its inverse 
ANG2PIX\_NEST is a convenient way to handle sets of rays with an adaptive angular density (see below).

\begin{figure}
\centering
\includegraphics[trim=0cm 1.5cm 0cm 0cm,scale=0.7,clip=true]{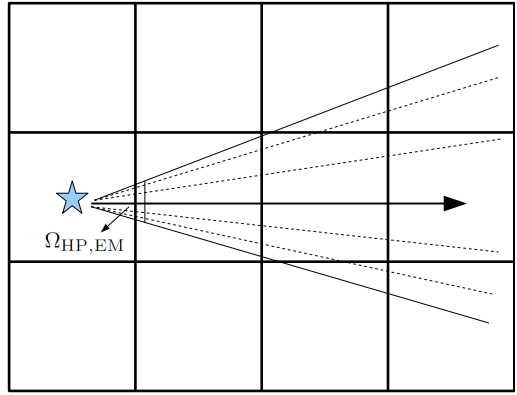}
\caption{Ray beam associated with an HEALPix pixel, propagating throughout the model. During the RT calculation, the variation
of $I_{\lambda}$ is followed only along the main direction (bold arrow). The same variation is assumed for the other directions within the 
same ray beam (dashed lines). Note that at large distances from emitting cell, the beam begins intersecting more cells simultaneously.}
\label{beam_prop}
\end{figure}

For each ray, the following approximations are implemented: \\
1) the calculation is performed as if all the cell luminosity propagates through solid angles defined by the pixels of an HEALPix sphere 
centred on the cell centre;\\
2) since one can only follow the exact variation of $I_\lambda$ along the finite set of HEALPix directions, it is assumed that the 
evolution of $I_\lambda$ along all the directions included within the total solid angle associated with a ray (determined by the adopted 
HEALPix scheme angular resolution) is exactly the 
same as for the main central ray direction (see Fig.\ref{beam_prop}).\\

The above approximations imply that the total luminosity density associated with a ray and flowing through an HEALPix solid 
angle $\Omega_{\rm{HP,EM}}$ 
at any distance $r$ from the centre of the emitting cell is given by:
\begin{equation}
 L_{\lambda,\rm{ray}}(r)=I_\lambda(r)\Omega_{\rm{HP,EM}}A_{EM}
 \label{lum_ray_beam}
\end{equation}
where $A_{\rm{EM}}$ is the projected area of the emitting cell (assumed to be equal to the emitting cell size squared). \\
The variation of $I_\lambda$, when the associated ray crosses a cell, is calculated using the following 
expressions. The variation due to the crossed optical depth $\tau_\lambda=k_\lambda\rho_c\Delta r$ and cell volume emissivity $j_{\lambda,c}$ 
within the cell originating 
the ray is equal to:

\begin{equation}
 I_{\lambda,0}=\frac{j_{\lambda,c}\Delta r}{\tau_\lambda}(1-e^{-\tau_\lambda})
\end{equation}
if $\tau_\lambda > 0$ (cell containing dust)\footnote{Actually the numerical implementation requires to assume a small threshold value higher than 
zero, such that $e^{-\tau_\lambda}\neq 1$ when the exponential function is evaluated.}. If $\tau_\lambda=0$, we assumed:
\begin{equation}
 I_{\lambda,0}=j_{\lambda,c}\Delta r
\end{equation}
which is consistent with the previous expression when $\tau_\lambda\rightarrow0$. During the scattering iterations, the scattered light is 
considered for the calculation of the cell volume emissivity, as it will be shown later. Instead, the new value of $I_\lambda$ after 
the crossing of a ray through a cell different from the original one is simply given by: 

\begin{equation}
 I_{\lambda,i+1}=I_{\lambda,i}e^{-\tau_\lambda}
\end{equation}

Apart from calculating a new value for $I_\lambda$, when a ray crosses a cell, the code determines the contribution of the ray to the intersected cell RFED
and the amount of energy scattered by the dust in the intersected cell. However, these contributions can be accurately calculated only if a sufficiently 
high ray angular density is used. In fact, depending on the cell sizes and the adopted HEALPix angular resolution, 
at any distance from the emitting cell, the ray beam as a whole (not just the main ray direction) can intersect either a single cell or a group of cells at 
the same time. At distances large enough from the emitting cell, the beam, originally intersecting only one cell at a time, will begin intersecting more 
cells simultaneously (see Fig.\ref{beam_prop}). As said before, the code traces the evolution of $I_\lambda$ only along the main direction of the ray beam and it is 
assumed that for all the directions within the beam the $I_\lambda$ variation is exactly the same. However, far away from the emitting cell, the beam propagates 
through larger physical volumes and the previous approximation can become very inaccurate. Actually, in order to obtain a precise calculation of the RFED 
contributed by an emitting cell $C_{\rm{EM}}$ to a certain cell $C_{\rm{INT}}$, it is desirable that several rays originating from $C_{\rm{EM}}$ are 
intersecting $C_{\rm{INT}}$. In this way, the calculation of RFED is more accurate because of the better estimation of the average $I_\lambda$ within 
the intersected cell. 

\begin{figure}
\centering
\includegraphics[trim=0cm 1.5cm 0cm 0cm,scale=0.8,clip=true]{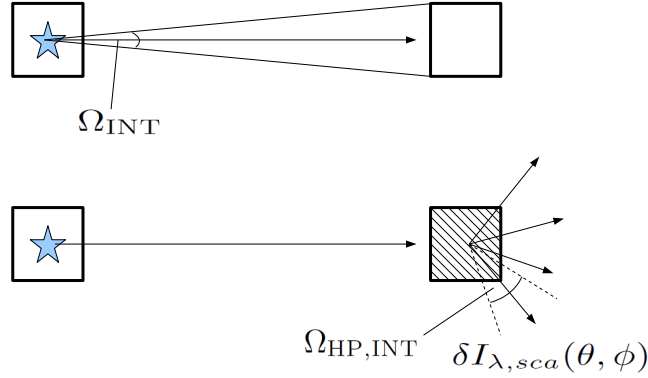}
 \caption{Definitions of the solid angle subtended by an intersected cell $\Omega_{\rm{INT}}$, the scattered intensity $I_{\lambda,\rm{sca}}(\theta,\phi)$ 
 from an intersected cell and the intersected cell HEALPix solid angle $\Omega_{\rm{HP,INT}}$.}
\label{i_sca}
\end{figure}

If one defines $A_{\rm{INT}}$, the projected area of the intersected cell (approximated by the cell size squared), and  
 $\Omega_{\rm{INT}}=\frac{A_{\rm{INT}}}{r^2}$, the solid angle subtended by the projected area of the intersected cell and with origin in the centre of 
the emitting cell (see Fig.\ref{i_sca}), in order to have more rays from $C_{\rm{EM}}$ crossing $C_{\rm{INT}}$ one requires that: 

\begin{equation}
\Omega_{\rm{HP,EM}} < \frac{\Omega_{\rm{INT}}}{N_{\rm{rays}}}
\label{beam_cond}
\end{equation}
where $N_{\rm{rays}}$ is an input parameter equal to the minimum number of rays which should cross $C_{\rm{INT}}$. 
If this condition is fulfilled by all the intersected cells 
within a certain region
 around an emitting cell, we say that the rays are ``fully sampling'' that region, using the same terminology already 
introduced in \S2.2.   

Once the above condition is fulfilled for an intersected cell, the code defines $\Omega=\Omega_{\rm{HP,EM}}$ and 
the contribution of a ray to the intersected cell 
RFED $\delta U_{\lambda,\rm{INT}}$ is given by 

\begin{equation}
\delta U_{\lambda,\rm{INT}}=\frac{<I_{\lambda}>A_{\rm{EM}}\Omega\frac{\Delta r}{c}}{V_{\rm{INT}}}
\label{en_dens_contr}
\end{equation}
where $c$ is the speed of light, $\frac{\Delta r}{c}$ is equal to the time needed by the light to cross the intersected cell,  
$V_{\rm{INT}}$ is the 
volume of the intersected cell and the average value of $I_{\lambda}$ along the crossing path is equal to: 
\begin{equation}
 <I_{\lambda}> = \frac{I_{\lambda,i}(1-e^{-\tau_\lambda})}{\tau_\lambda}  
\end{equation}
if $\tau_\lambda > 0$ and 
\begin{equation}
<I_{\lambda}> = I_{\lambda,i} 
\end{equation}
if $\tau_\lambda=0$. 

The ray contribution to the specific intensity scattered by the intersected cell $\delta I_{\lambda,\rm{sca}}(\theta,\phi)$ is 
(see Fig.\ref{i_sca}):

\begin{equation}
\delta I_{\lambda,\rm{sca}}(\theta,\phi)=
\frac{I_{\lambda,\rm{ext}}w_{\rm{\lambda,\rm{sca}}}A_{\rm{EM}}\Omega \Phi(\theta,\phi)}{A_{\rm{INT}}\Omega_{\rm{HP,INT}}}
\label{scatt_contr}
\end{equation}
where $\omega_{\rm{\lambda}}=\frac{k_{\lambda,\rm{sca}}}{k_{\lambda,\rm{ext}}}$, $\Omega_{\rm{HP,INT}}$ is the solid angle determined by the HEALPix angular resolution
adopted to store the scattered radiation intensity in the intersected cell, $\Phi_{HG}(\theta,\phi)$ is a term representing 
the integration of the Henyey-Greenstein scattering phase function over the solid angle $\Omega_{\rm{HP,INT}}$:

\begin{equation}
\Phi_{HG}(\theta,\phi)=\frac{\Omega_{\rm{HP,INT}}}{4\pi}\frac{1-g_\lambda^2}{[1+g_\lambda^2-2g_\lambda\cos (\theta)]^{3/2}]}
\label{HG_phase}
\end{equation}
and $I_{\lambda,\rm{ext}}$ is the amount of ray specific intensity absorbed or scattered by the cell, which is given by:

\begin{equation}
 I_{\lambda,ext}=I_{\lambda,i}\left(1-e^{-\tau_\lambda}\right)
\end{equation}

The scattered luminosity values (numerator of Eq. \ref{scatt_contr}) are accumulated for each cell on an 
array SCATT\_EN$(\theta,\phi)$ for a certain set of HEALPix directions, 
whose angular density 
can be specified in the input according to memory availability \footnote{The SCATT\_EN array can be very large: dimension = number of cells $\times$ 
$4\pi/\Omega_{\rm{HP,INT}}$.}. Of course, a higher numerical accuracy will be achieved if the scattered intensity values are stored
with a higher angular resolution. 

In case the condition expressed by Eq. \ref{beam_cond} is not fulfilled, all the equations above are still used but 
with $\Omega=\Omega_{\rm{HP,EM}}$ 
if $\Omega_{\rm{HP,EM}} < \Omega_{\rm{INT}}$ and $\Omega=\Omega_{\rm{INT}}$ if $\Omega_{\rm{HP,EM}} > \Omega_{\rm{INT}}$. That is, we consider only the 
beam luminosity passing through the intersected cell. As it will be explained later, condition \ref{beam_cond} is not required to be fulfilled
beyond the regions where full sampling is desired. That means, beyond a certain distance from an emitting cell there could be cells 
which are not 
intersected by any ray and therefore none of the above quantities is calculated by the code for those cells.        

The formulae given above are different when the crossed cell coincides with the emitting cell originating the rays. In this case
one has also to take into account the internal volume emissivity of the cell. Therefore,  
the average value of $I_\lambda$ needed to calculate the ray contribution to the local RFED by using Eq. \ref{en_dens_contr} is 
given by:

\begin{equation}
 <I_{\lambda}>=\frac{j_{\lambda,c}\Delta r}{\tau_\lambda^2}(e^{-\tau_\lambda}-\tau_\lambda-1)
\end{equation}
if $\tau_\lambda>0$ and

\begin{equation}
 <I_{\lambda}>=\frac{j_{\lambda,c}\Delta r}{2}
\end{equation}
if $\tau_\lambda=0$. Instead the value of $I_{\lambda,\rm{ext}}$ in Eq. \ref{scatt_contr} is given by:

\begin{equation}
I_{\lambda,\rm{ext}}=\frac{j_{\lambda,c}\Delta r\left(e^{-\tau_\lambda}+\tau_\lambda-1\right)}{\tau_\lambda}
\end{equation}
and all the quantities with subscript $\rm{INT}$ in Eq. \ref{en_dens_contr} and \ref{scatt_contr} 
refer to the emitting cell in this case. During scattering iterations, the term $j_{\lambda,c}\Delta r$ in the previous equations is substituted 
by $I_{\lambda,\rm{sca}}(\theta, \phi)$, the specific intensity of the total scattered radiation accumulated locally on each cell 
(see Eq.\ref{scatt_contr}).   

\subsection{The radiative transfer algorithm}

\subsubsection{Step 1: Calculation of a Lower Limit for the RFED}
\begin{figure}
\centering
\includegraphics[trim= 0cm 6cm 0cm 1cm, scale=0.5,clip=true]{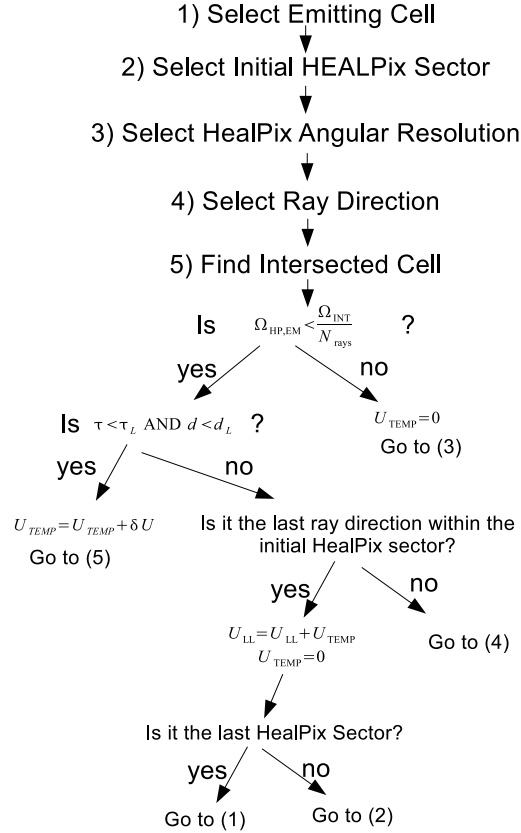}
\caption{Flow diagram of the procedure used to estimate the lower limit of the RFED.}
\label{flow_diag_fe}
\end{figure}

In the first step of the RT algorithm, ray-tracing is performed from each emitting cell by fully sampling 
all the cells until the rays cross a limit optical depth 
or distance specified in the input (see Fig.\ref{fe_ray_prop}). In this way the contributions from each emitting cell to the RFED in the 
areas around those cells are summed up to obtain
a lower limit $U_{\lambda,\rm{LL}}$ for the RFED distribution in the entire model. Full sampling of a region of cells centred on an emitting 
cell requires a sufficient amount of rays to be launched from the emitting cell. However, the ray angular density necessary to achieve full 
sampling can vary depending on the angular direction. In order to optimize the number of rays, our code has the additional peculiarity 
 that the angular distribution of rays originating from emitting cells is not homogeneous. 
 Specifically, the ray angular density is optimized within an initial HEALPix sector by the following iterative 
procedure (see flow diagram in Fig.\ref{flow_diag_fe}). 
 
A first ray-tracing attempt is performed with an initial HEALPix angular 
resolution, e.g. the calculation is performed only for the direction passing by the centre of an HEALpix main sector\footnote{see upper-left panel 
of Fig.\ref{hpix_sphere} for definition of ``HEALPix main sector''}.  
The beam defined by the chosen HEALPix pixel will intersect the cells progressively more distant from the emitting cell. As said in the previous subsection, 
in order to have an accurate estimation
of the radiation field contribution $\delta U_\lambda$ carried by the beam to an intersected cell, it is necessary that the entire beam is passing through the cell, together 
with at least several other adjacent beams. Therefore, when an intersected cell is found, the code checks if the condition expressed by Eq.\ref{beam_cond}
is fulfilled. When this happens, the contribution to the RFED is stored in a temporary array $U_{\rm{TEMP}}$ 
and the ray-tracing continues to the next intersected cell in the same way. When the above condition is not realized, the $U_{\rm{TEMP}}$ array is 
initialized 
and the radial ray-tracing re-starts from the beginning but with a higher HEALpix angular resolution. That is, more rays are launched 
within the initial HEALpix sector. In this way, smaller beams are generated 
until condition (\ref{beam_cond}) is always fulfilled for all the intersected cells within the input-defined distance or optical depth crossed by each ray. 
When this is realized, the ray-tracing can be performed without interruptions for all the rays within the chosen initial HEALPix sector. 
After the last ray within the initial HEALPix
sector has been processed, 
 the values of RFED stored in $U_{\rm{TEMP}}$ are added to the array $U_{LL}$ and then $U_{\rm{TEMP}}$ is initialized. This 
 iterative procedure is performed for all the initial HEALpix sectors, covering the entire sphere, and for each emitting cell. 
   
\subsubsection{Step 2: Processing Direct Radiation}
\begin{figure}
\centering
\includegraphics[trim= 0cm 2cm 0cm 0cm, scale=0.5,clip=true]{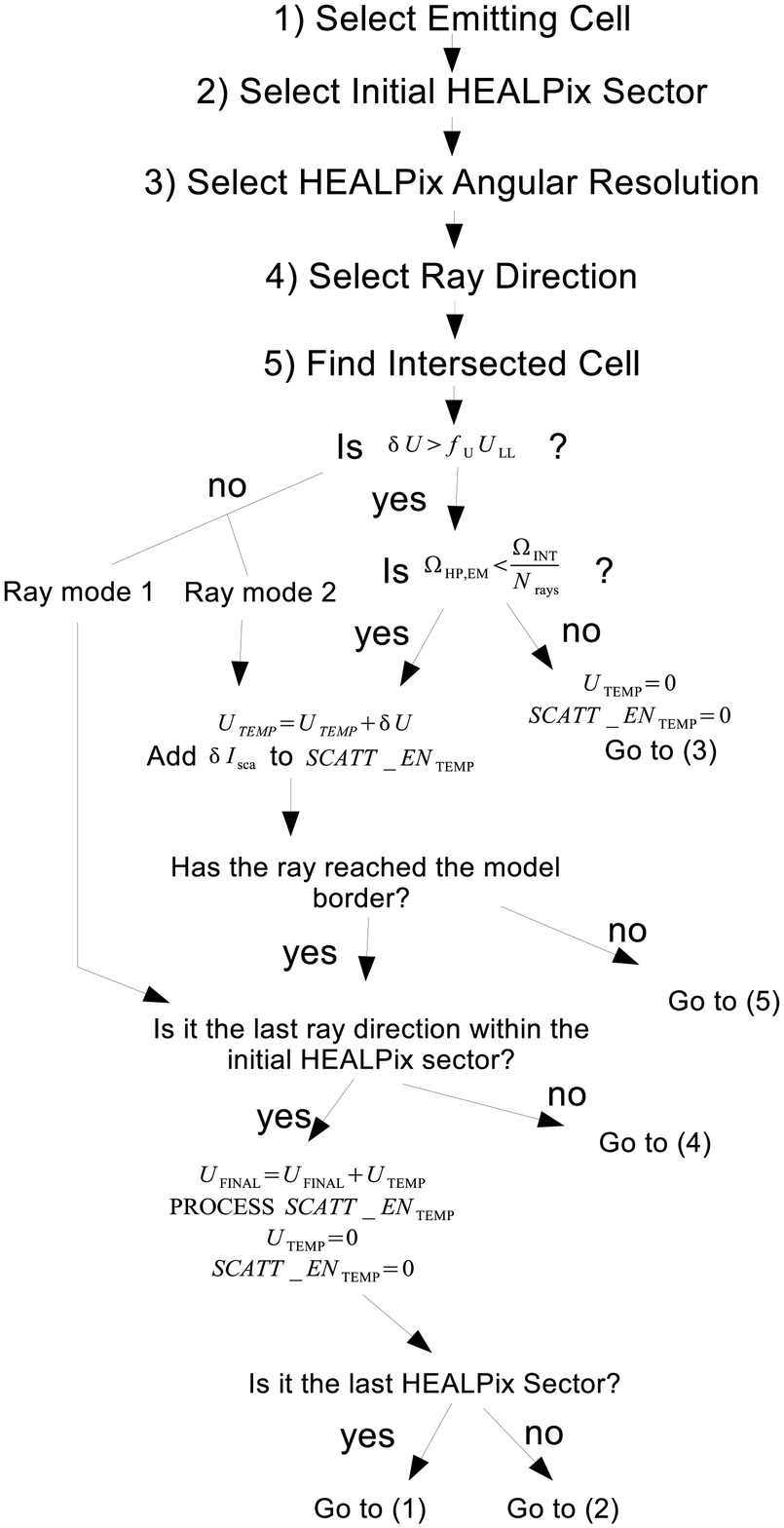}
\caption{Flow diagram of the main RT procedure (used for processing both the direct and scattered light).}
\label{flow_diag_rt}
\end{figure}

In the second step, the ray-tracing is started from the beginning again following the same iterative procedure explained in 
the previous subsection but with some differences (see flow diagram in Fig.\ref{flow_diag_rt}). 
As before, the angular density of the rays is increased until the 
intersected cells, where the 
contribution to the local RFED needs to be accurately calculated, are crossed by multiple rays. However, during this step
the increasing of ray angular density is stopped if the contribution by a ray to an intersected cell RFED is a very small 
fraction of the value stored at 
that position in $U_{\lambda,\rm{LL}}$. That is, when:
\begin{equation}
 \delta U_\lambda < f_U \times U_{\lambda,LL}
 \label{en_contr_cond}
\end{equation}
with $f_U$ defined in the input (see \S4). 
When this condition is fulfilled, it means that the emitting cell, originating the ray, is not contributing substantially to the RFED 
of the intersected 
cell under consideration and to all the other cells beyond that in the direction of the ray. In addition, at variance with step one,
ray-tracing is not necessarily limited in the region where full sampling is required, but optionally rays can keep propagating until the model 
border (this option will be referred to as ``ray mode 2''). 
Similarly as before, the values of the RFED contributions are 
first stored in $U_{\rm{TEMP}}$ and then 
added to the array $U_{\lambda,\rm{FINAL}}$, after the ray-tracing from all the directions within the initial HEALPix sector is completed. 
After this procedure has been performed 
for all the emitting cells, $U_{\lambda,\rm{FINAL}}$ contains all the contributions from direct radiation but still lacks the contribution from 
scattered light, which will be calculated in the next step. However, during this step, the scattered 
radiation energy information is stored after each cell crossing in the array SCATT\_EN($\theta,\phi$). 
As explained in section \S3.2, this array contains the scattered radiation 
luminosity in a 
finite number of HEALPix beams, whose angular density is defined by the user according to memory availability. 
When a ray crosses a cell, the fraction of 
scattered luminosity is angularly distributed according to the Henyey-Greenstein phase function 
(see Eq.\ref{HG_phase}).  
 
\subsubsection{Step 3: Processing Scattered Radiation}     
The last step consists of the processing of the scattered radiation. Since the scattered radiation itself can be scattered multiple 
times, this step requires
several iterations. However, the procedure for the ray-tracing is completely the same as the one in the previous step once the 
scattered luminosity values, 
stored in the SCATT\_EN array, are 
considered to calculate the volume emissivity from each cell. The only difference is that, while in the previous steps the emission from each
cell was 
isotropic, in this step the 
ray luminosities can depend on the angular direction. The SCATT\_EN values from each cell are 
on turn processed and initialized. 
Using the same optimisation procedure described above, ray-tracing from each cell is performed to calculate 
the contribution to $U_{\lambda,\rm{FINAL}}$ and to the SCATT\_EN arrays of the intersected cells. After all the emitting cells 
have been processed once, 
a check on the remaining luminosity $\Delta L_{\lambda}$ stored in the SCATT\_EN array is performed. If this luminosity is higher than a 
very small 
fraction $f_L$ of the total 
luminosity $L_\lambda$ emitted by the entire model, that is, when $\Delta L_{\lambda} > f_L L_{\lambda}$, a new scattering iteration 
is started. If not, scattering iterations are stopped and the output of the entire calculation is printed on a file.
The output includes both the RFED distribution and the escaping radiation specific intensity in several directions, 
calculated as described in the next section.

\subsection{Calculations of the Escaping Radiation Specific Intensity}
Although the algorithm we developed is optimized for the calculation of RFED distributions, our code can be used to derive the 
specific intensity of the radiation emitted or scattered by each cell and escaping outside the model volume. The code can 
calculate either averages for the radiation propagating within large solid angles or the specific intensity for the radiation propagating into 
single directions defined in the input. 
As it has been shown before, our code optimizes the angular density of the rays departing by each emitting 
cell in order to obtain a full sampling of cells where the ray RFED contribution is important. 
Beyond the region fully sampled, if specified in the input (so-called ``ray mode 2''), rays can 
keep propagating throughout the model until the model border (although they can miss a progressively higher fraction of cells). 
When a ray arrives to the model border, it is possible to calculate the specific intensity of the escaping radiation, 
generated by the emitting cell in the ray direction:

\begin{equation}
 I_{\lambda,\rm{esc}}=I_{\lambda,o} e^{-\tau_\lambda}
 \label{i_esc_formula}
\end{equation}
where $\tau_\lambda$ is the total optical depth crossed by the ray from the emitting cell to the model border. 
The calculation of $I_{\lambda,\rm{esc}}$  can be done for all the rays belonging to the sets of rays within an HEALPix main sector, within which
the ray angular density has been optimized as described in \S3.3. It is then straightforward to calculate the following average: 

\begin{equation}
 <I_{\lambda,\rm{esc}}>=\frac{\sum I_{\lambda,esc,i}\Omega_{HP,EM,i}}{\Omega_{\rm{HP,MS}}}
\end{equation}
where $\Omega_{\rm{HP,MS}}=4\pi/12$ is the solid angle of an HEALPix main sector and the sum is performed for all rays passing 
within that solid angle. The average value $<I_{\lambda,\rm{esc}}>$ derived in this way 
can be 
used to measure the escaping luminosity within a HEALPix main sector using Eq. \ref{lum_ray_beam} with $I_\lambda(r)=<I_{\lambda,esc}>$.
Note that this estimate assumes that the missed cells in the volumes beyond the fully 
sampled region have attenuation properties similar 
to those of the intersected cells at the same distance from the emitting cell. 

The estimate of the escaping luminosity within an entire 
beam is useful to calculate the total amount of stellar luminosity escaping from the system. However, one would 
also like to store the escaping radiation specific intensity in a set of directions, which is what it would actually be 
observed on astronomical maps. 
To do this, we store the values of escaping radiation specific intensity from each cell, calculated using formula (\ref{i_esc_formula}),
for a user defined set of directions.   
Since far away from the model the rays reaching the observer are all parallel, one can consider the escaping radiation 
specific intensity coming from each emitting cell along parallel directions in order to create visual maps of the model as 
seen from different view angles (see an example in Fig. \ref{spiral_model_comparison}).

\section{Notes on Approximations, Implementation and Performance}

3D dust RT requires extremely high computational resources if one wants to obtain numerically accurate solutions for 
systems where both radiation sources and dust are distributed over different spatial scales.  Practically, this implies that 3D dust RT 
codes need to be developed trying to balance the need for numerical accuracy with approximations reducing the total computational time 
to acceptable amounts.\\

While implementing the algorithm described in the previous section, we made use of the following approximations:\\
\\
1) the dust density value is assumed to be constant within each cell and, thus, within each cell crossing path. 
This allows us to use the simple exponential expression 
given by Eq.\ref{i_new_equ} to calculate the new value 
of the specific intensity of a ray after a cell crossing. However, the accuracy of the results is then affected by the spatial 
resolution of the grid. A more precise method would consist in using adaptive steps along a ray and calculating new values for the dust 
density at each position within the crossing path (see e.g. Steinacker et al. 2003). However, these extra calculations would increase 
substantially the total calculation time. \\
\\
2) the angular sampling of scattered light in each cell is performed using a finite number of solid angles covering the entire sphere
(see \S3 for more details). That is, the contribution to the 
scattered light, calculated after a 
cell intersection, is stored in a number of directions corresponding to the dimension of the local SCATT\_EN array, defined in 
\S3 (typically 48 or 96 directions are used in the calculation presented below). This means, that some of the information on the angular distribution of the scattered light is 
neglected because of the relatively coarse sampling on the sphere. \\
\\
3) the increase of angular density for the rays departing from each emitting cell is stopped when the condition expressed 
by Eq.\ref{en_contr_cond} for the ray contribution to the local RFED is realized. This is one of the characteristic features of our 
RT algorithm and 
probably the main one which allows us to reduce substantially the computational time. As mentioned in \S3.3 and \S3.4, 
the code we developed gives the possibility 
of either simply stopping the rays at the locations where the above condition is realized (ray mode 1) or only avoiding the further increase of 
ray angular density beyond those points but still continuing the ray-tracing until the model border (ray mode 2). In both cases, 
a fraction of the luminosity carried by the beam associated with the rays is not considered 
in the calculation of the RFED distribution. Controlling the actual amount of this ``lost luminosity'', not processed during 
the RT calculation, is fundamental to assure 
that approximate global energy balance is reached between the luminosity emitted by the stars and the sum of the luminosities
absorbed and escaping from the system. In order to check this, we set up a counter 
of ``lost luminosity'' which sum up the ray luminosity which is not processed during the calculation. Contributions to this sum are 
provided by the entire beam luminosity at the locations where condition \ref{en_contr_cond} is realized. 
The input parameters directly affecting 
the amount of ``lost luminosity'' are $f_U$ and $N_{\rm rays}$, which have to be chosen carefully such to guarantee energy balance. 
Practically, one first finds an appropriate combination of $f_{U}$ and $N_{\rm rays}$ for a typical model such to obtain a low fraction
of ``lost luminosity'' at the end of the calculation. After that, it is usually enough to use the same combination for these parameters 
for similar models
and check that approximate energy balance has been achieved after each calculation. 
For all the calculations presented in the next section, we checked that, for the 
assumed combinations of those input parameters, the total lost luminosity is always less than 1-2\% of the total stellar luminosity. \\

The algorithm has been implemented in a \textsc{Fortran 90} code, which is a typical language used in high-performance computing (HPC).
Parallelization of the code has been 
implemented using the application
programming interface OpenMP, which allows parallel computing on shared-memory machines. We opted for OpenMP parallelization
since it required only a relatively small number
of extra lines within the existing code. Furthermore, given the additive nature of the RT problem, it has been possible 
to develop simple programs to distribute the 
calculations for different sets of emitting cells among different nodes in a computer cluster. Within each node the calculation can be 
performed in parallel 
mode without the need of communication between nodes, except when arrays sums are needed at the end of each step of the RT algorithm or 
between scattering iterations. Extra routines have been written to perform these sums and this has been sufficient to perform RT calculations
on computer clusters without the need of more elaborated MPI parallelization.\\    

The test runs presented in the next subsection have been 
performed using the computing facilities at the University of Central Lancashire in Preston (in particular the local HPC cluster) 
and at the Max Planck Institute f\"{u}r Kernphysik in Heidelberg. The computational times and memory required vary depending on the geometry 
of the model, the spatial resolution adopted and the number of CPUs used. For example, a typical single wavelength calculation 
for a disc galaxy model (see \S5.2) with a grid containing of the order of $10^5$ cells can take about 2 days by using 64 CPUs with 2.57 GHz 
clock rate (and by using a $f_U$ parameter low enough to achieve energy balance within less than a percent). In terms of memory, this
calculation requires few gigabytes of RAM memory (exact amount varies during the calculation because of the temporary ``allocatable'' 
arrays vastly used by our code).  
The mentioned calculation time seems to 
be rather longer compared to those needed by  
MC codes where acceleration techniques are well developed and which are able to handle multiwavelength photon packages 
simultaneously (see e.g. Baes et al. 2011, Jonsson 2006). However, one should notice that, at variance with ray-tracing codes, 
MC codes are not typically used  
to obtain accurate calculations of the monochromatic RFED distribution. Applications of MC methods focus on obtaining
a good calculation for the escaping multiwavelength spectra. In order to predict the dust emission, this requires only the dust 
temperature in each cell to converge and not necessarily each single value of the RFED at each 
wavelength.

\section{Comparisons with other codes}
In the following we describe and show the results of the comparisons with two codes: the 1D code \textsc{DUSTY} (\S5.1) and the 2D code used in 
Popescu et al. (2011) (\S5.2)

\subsection{Comparison with \textsc{DUSTY} code}

We performed a series of tests by comparing the results provided by our code with 
those obtained using the latest version of the RT code \textsc{DUSTY} (Ivezic \& Elitzur 1997). Specifically, we considered the geometric configurations 
of the benchmark solutions of Ivezic et 
al. (1997, I97), with parameters equivalent or similar to those adopted in that paper. 
The \textsc{DUSTY} code can be used to solve 
the dust RT problem for a geometry consisting of a single central radiation source illuminating a spherically symmetric 
shell of dust with 
an arbitrary dust density radial profile. In this case, one can reduce the 3D RT equation to 
a 1D equation for the mean radiation field intensity by taking advantage of the spherical symmetry of this 
configuration. Furthermore, the \textsc{DUSTY} code 
utilizes several scaling 
properties of the RT problem. These are used to calculate sets of solutions which do not depend on the absolute values 
of the luminosity of the 
central source and of the dust density and opacity but only on their spatial variation or wavelength dependence. Thus, given a dust 
radial density profile 
in the shell and the wavelength dependence of the dust optical properties, this specific RT problem can be completely defined 
once the following 
parameters are specified: \\ 
- the effective temperature $T_{\rm source}$ of the central source, which we assumed it emits as a black body  \\ 
- the dust temperature $T_1$ at the inner radius of the shell \\
- the total radial optical depth $\tau$ \\ 
- the outer to inner radius ratio $Y=r_2/r_1$. \\
     
As in I97, we performed RT calculations for two types of radial profiles of the dust density distribution: \\

1) a constant density profile $\rho(r)=\rho_o$; \\

2) a power-law density profile $\rho(r)=\rho_o\left(\frac{r}{r_1}\right) ^{-2}$\\ 

for $r$ in the range $[r_1,r_2]$ and $\rho(r)=0$ for any other value of $r$. \\

The dust opacity coefficients per unit dust mass are defined as follows: \\
\begin{equation}
  q_{\lambda,abs}=q_{\lambda,\rm{sca}}=1
\label{q1}   
\end{equation}
for $\lambda < 1\mu m$ and 
 
\begin{equation}
q_{\lambda,abs}=\frac{1}{\lambda} \; ; \; q_{\lambda,\rm{sca}}=\frac{1}{\lambda^4} 
\label{q2}
\end{equation}
for $\lambda > 1\mu m$.\\

Scattering is considered isotropic in the \textsc{DUSTY} code, which corresponds to assuming $g_{\lambda}=0$ in the scattering 
Henyey-Greenstein phase function used by our code.  \\ 

For both forms of dust density profiles specified above, we performed two series of tests where we compared the results for the dust 
temperature
radial profile and the outgoing radiation spectra. First, we created a grid of models by varying the inner radius dust 
temperature $T_1$ and keeping all the other parameters fixed to the 
same values. For a fixed amount of dust mass, a higher value of $T_1$ implies a higher dust luminosity which can be self-absorbed by dust. 
Since self-absorption is not included in our code yet, at least some of the discrepancies evidenced by this test can be due to this effect. 
We will refer to this test in the following as ``the dust temperature'' test. In a second series of tests, 
we have varied the value of $\tau$ while 
maintaining all the other parameters constant. By increasing the value of $\tau$, the source luminosity is absorbed more efficiently and
 dust heating due to self-absorption could become progressively more dominant, especially at larger radii. 
We will refer to this test as ``the optical depth'' test. \\   

The specific parameters used in the ``dust temperature'' test are the following: Y=1000, $T_{\rm source}=2500$ K, optical depth at 
1$\mu$m $\tau_1=1$ 
and $T_1=200, 400, 800$\,K. In the ``optical depth'' test we used these parameters: Y=1000, $T_{\rm source}=2500$ K, $\tau_1=2,5,10$ 
and $T_1=200$\,K. \\

The first step performed by our code is the creation of a spatial grid sampling the entire model.
To do this we require the absolute values of the central source luminosity and the physical distances 
corresponding to the inner and outer 
radius of the shell $r_1$ and $r_2$, as these quantities are not explicitly specified in the input parameters of the \textsc{DUSTY} code. 
We derived these quantities from the standard output of the \textsc{DUSTY} code, which assumes that the 
bolometric central source 
luminosity is equal to $10^4~\rm{L}_\odot$. Also, we obtained the absolute scaling of the 
dust density distribution $\rho_o$ by using the following formulae: 

\begin{equation}
\rho_o=\frac{\tau_1}{q_{1,\rm{ext}}(R_2-R_1)}
\end{equation} 
for the case of the constant dust density radial profile and  

\begin{equation}
\rho_o=\frac{\tau_1}{q_{1,\rm{ext}}\frac{R_1}{R_2}(R_2-R_1)}
\end{equation} 
for the case of the power law dust density radial profile. In the previous formulae $q_{1,\rm{ext}}=q_{1,abs}+q_{1,\rm{sca}}$ and the opacity 
coefficients are all evaluated at $1\mu m$. We used the coefficients at this wavelength since they are the highest. This implies that
the cell optical depths of the grid so created are the same or smaller at other wavelengths. For simplicity we used the same grid for all 
the wavelengths. \\

While creating the grid, we assigned average density values to each leaf cell containing dust. In the constant density profile case we 
simply assumed $<\rho_{\rm dust}(r)>=\rho_o$. In the power law case, we used the following expression for the cell dust density, 
corresponding to the cell density average along a radial direction: 

\begin{equation}
 <\rho_{\rm dust}(r)>=\rho_o R_1^2[\frac{1}{r_c-\Delta r/2}-\frac{1}{r_c+\Delta r/2}]/\Delta r
\end{equation}
where $r_c$ is the radius corresponding to the cell centre and $\Delta r$ is equal to the cell size. In the cases of the cells at 
the inner or outer 
border of the shell, we calculated averages by integrating the density only in the part of the cell containing dust and 
then dividing by the cell size. \\

We imposed the following conditions in the input of the grid creation program: 
1) a cell has to be subdivided in smaller cells if the cell optical depth exceeds a small fraction of the total radial optical depth
$\tau_1$, typically a factor of 0.01-0.03; 2) the minimum cell subdivision level is 3; 3) the maximum allowed cell subdivision 
level is equal to 7 and 8 for the constant and power law dust density profile, respectively; 4) in the case of the power law density 
profile, we also added the constraint 
that the maximum cell optical depth gradient is $\Delta\tau_{1\mu m}/\tau_{1\mu m}=0.4$. These conditions have been chosen in order to 
achieve a solution with good numerical accuracy but also to avoid to create too many cells (that is, less than $\approx10^{6}$ cells), 
thus  reducing the amount of calculation time needed. In fact, note that the amount of cell subdivisions is limited
by condition (3). Thus, in some parts of the model, a cell subdivision required by conditions (1) or (4) 
might not be performed because it would conflict with condition (3). The maximum and average values of the cell optical depths in each model 
are shown 
in Tables \ref{delta_tau_td_up} and \ref{delta_tau_tau_up}. \\

 \begin{table}
 \begin{center}
 \begin{tabular}{lllllll}
p & $\Delta\tau_{1,\rm{max}}$ & $<\Delta\tau_1>$ & $f_U$ & $N_{\rm{rays}}$ & $f_L$ & N$_{\rm cells}$ \\ \hline
 0 &  0.029 & 0.023 & 0.0001 & 16 & 0.001 & 302481\\
 2 & 0.23 & 0.006 & 0.001 & 16 & 0.001 &  214326 \\
 \end{tabular}
\end{center}
 \caption{3D grid and input parameters for the ``dust temperature'' test. Column 1) p=0 refers to the constant dust density case, 
p=2 to the power law case; Columns 2-3) maximum and average cell optical depth at 1$\mu$m; Column 4) ray energy contribution
threshold parameter; Column 5) minimum number of rays for cells within full sampling regions; Column 6) fraction of unprocessed total luminosity needed 
to end code iterations; Column 7) total number of cells. Note that the parameters do not depend on the specific value of $T_1=200,400,800$\,K. The total radial optical depth 
$\tau_1$ is always equal to 1 for all the models. } 
\label{delta_tau_td_up}
\end{table}

\begin{table}

 \begin{center}
 \begin{tabular}{llllllll}
p & $\tau_1$ & $\Delta\tau_{1,\rm{max}}$ & $<\Delta\tau_1>$ & $f_U$ & $N_{\rm{rays}}$ & $f_L$ & N$_{\rm cells}$ \\ \hline
 0 & 2 & 0.06 & 0.046 & 0.0001 & 16 & 0.001 & 302481 \\
 0 & 5 & 0.15 & 0.11 & 0.001 & 4 & 0.001 & 302481 \\
 0 & 10 & 0.29 & 0.23 & 0.001 & 4 & 0.001 & 302481\\
 2 & 2 & 0.45 & 0.013 & 0.001 & 16 & 0.001 & 214326\\
 2 & 5 & 1.13 & 0.032 & 0.001 & 16 & 0.001 & 214326\\
 2 & 10 & 2.2 & 0.064 & 0.001 & 16 & 0.001 & 214326\\ 
 \end{tabular}
\end{center}

 \caption{3D grid and input parameters for the ``optical depth'' test. Column 1) p=0 refers to the constant dust density case, 
p=2 to the power law case; Column 2) total radial optical depth at 1$\mu m$; Columns 3-4) maximum and average cell optical depth at 1$\mu$m; 
Column 5) ray energy contribution threshold parameter; Column 6) minimum number of rays for cells within full sampling regions; Column 7) fraction of unprocessed 
total luminosity needed to end code iterations; Column 8) total number of cells. Note that the inner dust temperature is always $T_1=200$K for all the models.} 
\label{delta_tau_tau_up}
\end{table}

Once the grid has been created, before starting the RT calculation, one has to assign values to the following parameters 
(see \S3): 
1) $f_U$, the relative energy density contribution threshold above which a ray contribution to a crossed cell RFED is 
considered non negligible; 
2) $N_{\rm{rays}}$, the minimum number of rays which has to cross a cell when the ray contributions are found not negligible;
3) $f_L$, the escaping luminosity threshold parameter, used to determine when the scattering iterations have to stop.
The adopted values are also shown in Tables \ref{delta_tau_td_up} and \ref{delta_tau_tau_up}. 

For each model, we performed the calculation at the following wavelengths: $0.443, 1.05, 1.1, 1.15, 1.2, 1.259, 1.3, 1.8, 2.2, 5.0, 10, 
20 \mu$m. 
We calculated only one point for $\lambda< 1\mu m$, that is at $\lambda=0.443\mu m$, because the opacity coefficients
used by I97 are the same in that wavelength range (see Eq. \ref{q1} and \ref{q2}) and the inferred RFED scales 
only with the luminosity of the central source at the different wavelengths. For $\lambda > 1\mu m$, the chosen wavelength steps are 
smaller between 1 and 2 $\mu m$. 
Since $T_{source}=2500 K$, the 
emission from the central source peaks in that wavelength region while the dust opacity is still relatively high. As a consequence,  
a consistent fraction of source luminosity is absorbed or scattered 
within the system at those wavelengths. Therefore, a good sampling in that wavelength region is desirable to obtain 
a more accurate solution for the dust temperature. At longer wavelengths, a finer sampling is not as important since both the opacity 
and the radiation intensity from the central source decrease rapidly. \\

After the RT calculation has been performed at all the wavelengths specified above, we obtained a grid of radiation 
field energy density 
spectra at each cell position. Then, we calculated the energy density values in the entire wavelength range $0.01-1\mu m$ by scaling 
the value 
inferred at $\lambda=0.443\mu m$ according to the central source luminosity at different wavelengths. This is possible because for $\lambda < 1 \mu m$ the 
dust opacity is assumed to be constant. In the wavelength range $1-100\mu m$ we simply 
interpolated the inferred 
values within that range. Then, in a way consistent with the \textsc{DUSTY} code calculation, 
we derived the equilibrium dust temperature $T_d$ at each position such that:

 \begin{equation}
  \int{q_{\lambda,\rm{abs}}B_{\lambda}(T_d)d\lambda}=\frac{c}{4\pi}\int{q_{\lambda,\rm{abs}}U_{\lambda}d\lambda}
 \label{loc_en_bal_dusty}
  \end{equation}

We also calculated the spectra of the outgoing radiation flux at the outer radius of the shell, a quantity given in the output of the 
\textsc{DUSTY} code, as follows. First, we derived the total escaping luminosity $L_{\lambda,\rm{esc}}$ by taking advantage of the cell average 
escaping brightness $<I_{\lambda,\rm{esc}}>$ within each HEALPix main direction, as provided by our code (see \S3.4). In fact, $L_{\lambda,\rm{esc}}$
can be expressed as: 

\begin{equation}
 L_{\lambda,\rm{esc}}=\sum_i <I_{\lambda,\rm{esc}}>_i \Omega_{\rm{HP,MS}}^iA_{\rm{EM}}^i
\end{equation}
where the sum is performed for all the leaf cells and all the HEALPix main sectors. The outgoing flux can then be derived by simply dividing
$L_{\lambda,\rm{esc}}$ by $4\pi R_2^2$. In order to obtain the values of $L_{\lambda,\rm{esc}}$ for a larger set of wavelengths, we used the 
same procedure already applied for the scaling and interpolation of the inferred RFED (see above).
We also calculated the contribution to the outgoing flux due to dust emission, which we derived 
assuming that the dust in each layer of the shell emits accordingly to the dust temperature radial profile we derived. Finally, we summed up 
both the central source and dust emission contributions to obtain the total outgoing radiation spectra.\\ 

\begin{figure}
\includegraphics[scale=0.7]{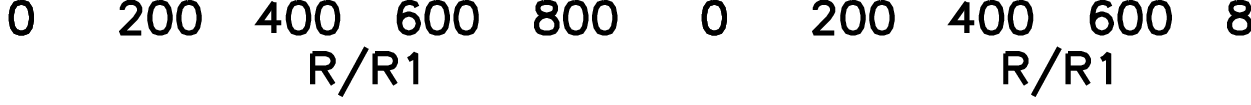}
\caption{Dust temperature radial profile in the constant dust density case and for the ``dust temperature'' tests. 
The plotted diamond symbols represent the temperature values
inferred by our code. The continuous line is the \textsc{DUSTY} code solution.}
\label{td_up_s2500_p0}
\end{figure}

\begin{figure}
\includegraphics[scale=0.7]{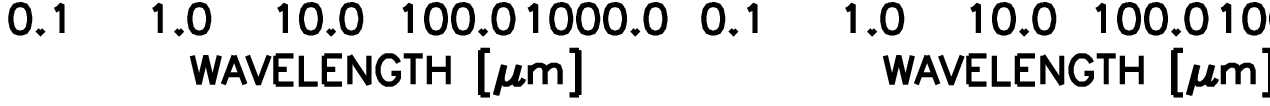}
\caption{Outgoing radiation flux in the constant dust density case and for the ``dust temperature'' tests. 
The plotted triangles represent the outgoing fluxes, for the radiation originating from the central source, inferred by our 
3D RT calculations (note that the plotted values do not include the dust emission). The diamonds represent the sum of 
the interpolation of the fluxes calculated by 
our code at different wavelengths (see text for details) 
plus the dust emission outgoing flux derived from the inferred dust temperature radial profile. The continuous line is the \textsc{DUSTY} code 
solution.}
\label{td_up_s2500_p0_flux}
\end{figure}

\begin{figure}
\includegraphics[scale=0.7]{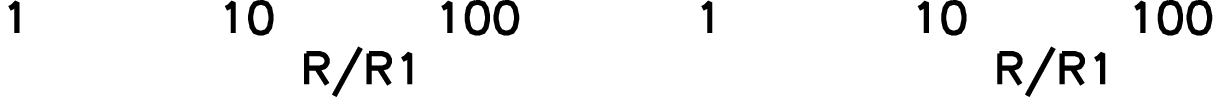}
\caption{Dust temperature radial profile in the case of the power law dust density profile and for the ``dust temperature'' tests. 
Same symbols as in Fig.\ref{td_up_s2500_p0}}
\label{td_up_s2500}
\end{figure}

\begin{figure}
\includegraphics[scale=0.7]{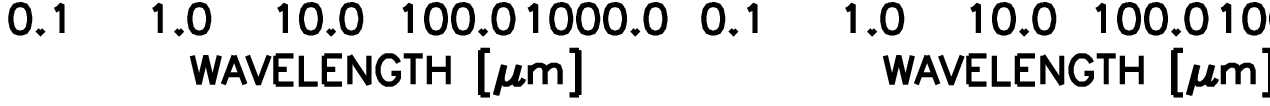}
\caption{Outgoing radiation spectra in the case of the power law dust density profile and for the ``dust temperature'' tests. Same symbols as 
in Fig.\ref{td_up_s2500_p0_flux}}
\label{td_up_s2500_flux}
\end{figure}

\begin{figure}
\includegraphics[scale=0.7]{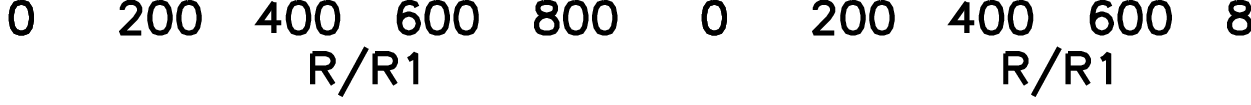}
\caption{Dust temperature radial profile in the constant dust density case and for the ``optical depth'' tests. Same symbols as in Fig.\ref{td_up_s2500_p0}}
\label{tau_up_s2500_p0}
\end{figure}

\begin{figure}
\includegraphics[scale=0.7]{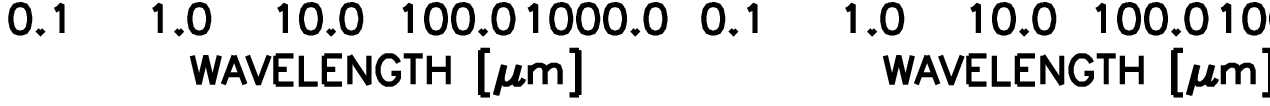}
\caption{Outgoing radiation spectra in the constant dust density case and for the ``optical depth'' tests. Same symbols as 
in Fig.\ref{td_up_s2500_p0_flux}}
\label{tau_up_s2500_p0_flux}
\end{figure}

\begin{figure*}
\includegraphics[scale=0.7]{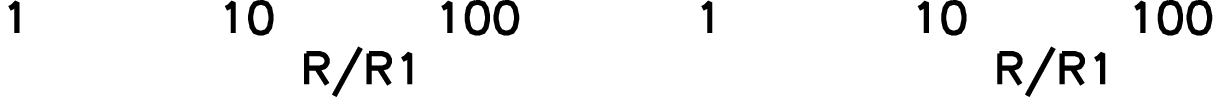}
\caption{Dust temperature radial profile in the case of the power law dust density profile and for the ``optical depth'' tests. Same symbols as in Fig.\ref{td_up_s2500_p0}}
\label{tau_up_s2500}
\end{figure*}

\begin{figure*}
\includegraphics[scale=0.7]{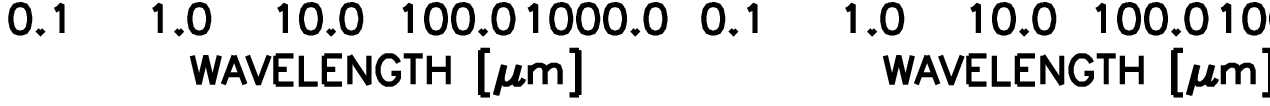}
\caption{Outgoing radiation spectra in the case of the power law dust density profile and for the ``optical depth'' tests. Same symbols as 
in Fig.\ref{td_up_s2500_p0_flux}}
\label{tau_up_s2500_flux}
\end{figure*}

In Fig.\ref{td_up_s2500_p0}-\ref{tau_up_s2500_flux} we show the comparison of the dust temperature radial profiles and outgoing 
radiation spectra we inferred with our code with those obtained by the \textsc{DUSTY} code. Fig.\ref{td_up_s2500_p0} and \ref{td_up_s2500_p0_flux}
shows the results for the constant density profile for the ``dust 
temperature'' test. As shown in Table \ref{sigma_ivezic_dust temp}, the average difference between the \textsc{DUSTY} code dust temperature profile and the one generated by our code is 
about $1-2\%$ for all the values of $T_1$. The average difference between the inferred outgoing spectra is about $2-3\%$ for the 
points actually calculated by our code (triangles in the figure, hereafter referred to as ``calculated fluxes''. In this 
 estimate of the discrepancy we did not consider the near-infrared points which are dominated by dust emission). The discrepancy increases
 to $5-7\%$ if one consider the entire UV-to-IR SED we derived as explained above. The highest 
discrepancies are observed in the MIR region. An excess of flux in the MIR range is expected since we did not take into account 
dust self-attenuation in our code and dust opacity is still relatively high at MIR wavelengths.    
For the power law density profile, the discrepancy between the inferred temperature profiles/SEDs increases while going to 
higher values of $T_1$, as shown in 
Fig.\ref{td_up_s2500} and \ref{td_up_s2500_flux}. Specifically, the average difference between the inferred dust temperature profiles is 
equal to 1.6\%, 2.3\% and 3.4\% for $T_1$ equal to 200, 400 and 800 K respectively. We note that the discrepancy is due to a systematic 
underestimation of the dust temperature by the 3D code, which is also expected because we did not include the extra-heating due to 
self-absorption. For the outgoing radiation spectra the average difference 
is in the range $2-4\% $ for the calculated fluxes and about $5-13\%$ for the global SEDs, with the highest discrepancies again in
the MIR range. \\
The results of these tests show that there is only a rather small difference for the dust temperature radial profile 
inferred by the two codes for models at fixed optical depth $\tau_1=1$ and with dust temperature $T_1$ varying between 200 and 800 K. 
However, the discrepancy is higher for the outgoing spectra. \\

Fig.\ref{tau_up_s2500_p0}-\ref{tau_up_s2500_flux} show the results for the ``optical depth'' tests for the constant and power law dust 
density profiles. Average discrepancies are tabulated in Table \ref{sigma_ivezic_opt_depth}. 
In the case of the constant density profile (see Figs.\ref{tau_up_s2500_p0}-\ref{tau_up_s2500_p0_flux}), 
all the three models ($T_1=200$ and $\tau_1=2,5$ and $10$) 
present an average difference for the dust temperature 
profiles of order of $2\%$. The outgoing radiation spectra differ on average by $5-17\%$ for the calculated fluxes and $9-29\%$ 
for the total SEDs.  For the power law profile, the average differences in the temperature profiles are 3.2\%, 6.5\% and 15\% going
from $\tau_1=2$ to $\tau_1=10$. Instead the 
differences for the outgoing radiation spectra are within $6-16\%$ for the calculated fluxes and $8-30\% $ for the total 
inferred SEDs. The discrepancy increases systematically with the optical depth of the model considered. \\ 

To summarize, from the comparison with the \textsc{DUSTY} code we obtained the following results. For the dust temperature test, we found that:\\ 
- the dust temperature radial profile and the calculated fluxes agree within few percent for both the constant and power law dust density profile\\
- the average discrepancy for the total SEDs is of order of $5-10\%$, with the highest discrepancies in the MIR region.  \\
For the optical depth test, we found that:\\
- the dust temperature radial profiles agree within few percent for the constant dust density case, while for the power law case 
the discrepancy increases with the optical depth of the model (up to 15\% for $\tau_1=10$)\\
- The discrepancy for both the calculated fluxes and the total SEDs increases with the model optical depth in a similar way for both the 
dust density profiles. For the highest optical depth $\tau_1=10$, it is of order of 15\% for the calculated fluxes and 30\% for the 
total SEDs. As before, the highest discrepancies when comparing the total SEDs are found in the MIR region.  \\

Although the results provided by the two codes seem to be quite consistent for models with optical depths $\tau_1=1$ and $2$, there is still 
some residual discrepancy for more optically thick models. A certainly important cause of the observed discrepancy is that, as already 
pointed out before, dust 
self-heating needs to be included in the code in order to predict accurate dust temperature profiles and output spectra, especially
in the MIR region. However, another source of error is also the relatively low resolution of the 3D calculation compared to the 1D one. 
As explained before, the 3D grid spatial resolution is higher in regions with higher dust density but it has been limited to keep the 
total number of cells in the range $10^5-10^6$. Thus, in the grids used in the 
calculations some
regions have relatively high optical depths (see maximum cell optical depths in Table \ref{delta_tau_tau_up}). 
An increased spatial resolution would have been beneficial 
to improve the accuracy of the solution but at the expense of a much longer computational time. This problem is more important 
 for models with higher optical depths and might explain the residual discrepancy for the calculated fluxes in the UV-optical regime.\\  

Because of the lack of dust self-heating in our code and the RT geometry assumed in this test, the solutions provided by the \textsc{DUSTY} code do 
not provide ideal benchmarks to test our code. In particular, the geometry of the emission source/opacity of a star/dust shell
 does not resemble that of a galaxy,
which is the class of object for which we developed our algorithm. For these reasons, we decided to compare solutions for a galaxy type 
geometry of stars and dust, using the 2D calculations of Popescu et al. (2011). The results of this comparison are shown in the next subsection.

\begin{table}
\begin{center}
\begin{tabular}{ccccc}
p & $T_1$  & $<\frac{\Delta T}{T}>$ & $<\frac{\Delta\lambda F_\lambda}{\lambda F_\lambda}>_{\rm{CALC}}$ & $<\frac{\Delta\lambda F_\lambda}{\lambda F_\lambda}>_{\rm{SED}}$ \\\hline 
0 & 200 K  & 0.018 & 0.025  & 0.07   \\
0 & 400 K  & 0.017 & 0.035   &  0.05   \\
0 & 800 K  & 0.01 & 0.03    & 0.068   \\
2 & 200 K  & 0.016 & 0.024 & 0.052  \\
2 & 400 K  & 0.023 & 0.03 & 0.063  \\
2 & 800 K  & 0.034 & 0.043 &  0.13  \\\hline 
\end{tabular}
\end{center}
 \caption{Average relative discrepancies for the models of the ``dust temperature test'': Col.1) p=0 for the constant density model, p=2 for the power law model; col.2) inner shell dust temperature,
 col.3-5) average relative discrepancy for the dust temperature radial profile, the calculated fluxes and the global SEDs (see text).} 
\label{sigma_ivezic_dust temp}
\end{table}

\begin{table}
\begin{center}
\begin{tabular}{ccccc}
p & $\tau_1$ & $<\frac{\Delta T}{T}>$ & $<\frac{\Delta\lambda F_\lambda}{\lambda F_\lambda}>_{\rm{CALC}}$ & $<\frac{\Delta\lambda F_\lambda}{\lambda F_\lambda}>_{\rm{SED}}$ \\\hline 
0 &  2 &  0.019 & 0.052 & 0.087  \\
0 &  5 &  0.022 & 0.14 & 0.2    \\ 
0 &  10 & 0.021 & 0.17 & 0.29  \\
2 &  2 & 0.032 &   0.064 & 0.08  \\
2 &  5 & 0.065 &  0.14 & 0.15 \\
2 &  10 & 0.15 & 0.16 & 0.3   \\

\end{tabular}
\end{center}
 \caption{Average relative discrepancies for the ``optical depth test'': Col.1) p=0 for the constant density model, p=2 for the power law 
density model; col.2) radial optical depth at 1$\mu$m,  col.3-5) average relative discrepancy for the dust temperature radial profile, 
the calculated fluxes and the global SEDs (see text).} 
\label{sigma_ivezic_opt_depth}
\end{table}

\subsection{Comparison with 2D calculations of Popescu et al. (2011)}

Most of the dust RT solutions considered as benchmarks in the literature (e.g. Ivezic et al. 1997, Pascucci et al. 2004) 
are designed to test RT codes in cases resembling star forming clouds or proto-planetary discs. Those systems can be well approximated 
by a central luminous source illuminating a spherical dust distribution or a dusty disc. Although able to handle completely arbitrary
geometries, our code has been 
designed for the purpose of solving the RT problem in a galaxy type geometry, which consists of an extended distribution of sources illuminating 
a dust distribution. For this reason, we decided to perform a comparison with the 2D RT calculations presented by 
Popescu et al. (2011, P11), which assume a disc galaxy geometry. The accuracy of the solutions for the radiation fields from P11 
have been tested in Popescu \& Tuffs 2013 against analytic solutions. For the cases where the analytic solution is an 
exact solution, the accuracy of the 2D code has been proven to be better than 1\%. 

P11 used a modified version of the Kylafis \& Bahcall (1987) 2D ray-tracing code to calculate radiation fields within a galaxy model 
comprising of three stellar components (a bulge, a disc and a thin disc) and two dust discs (called ``thick dust disc'' and ``thin dust disc'').
Both the stellar volume emissivity and dust density for the disc type components are described by a double exponential distribution:  

\begin{equation}
 f(R,z)=f(0,0)\exp^{-\frac{R}{h_{d,s}}-\frac{|z|}{z_{d,s}}}
 \label{double_exp}
\end{equation}
where $h_{d,s}$ and $z_{d,s}$ are the scale-length and scale-height of the disc components. As described in P11, the stellar discs 
(referred to as ``disc'' for the old stellar component and ``thin disc'' for the young stellar component) are characterized by the
geometrical parameters scale-length and -height, as reported in Table E1 of that paper. They are also described by two parameters 
$old$ and $SFR$. The $SFR$ is a parameter defining the luminosity of the young stellar population (the thin disc)
and $old$ is a parameter defining the luminosity of the old stellar population (the disc). The spectral luminosity densities 
corresponding to the unit values of these parameters, $SFR=1\,M_{\odot}/yr$ and $old=1$ are given in Table E2 of P11\footnote{We 
note that the models are truncated at the radius $R=24$\,kpc. The unit luminosities in Table E2 of P11 refer to the luminosities within
the truncation radius.}. The parameter values for the stellar discs in the B-band and the dust discs are also shown in 
Table \ref{table_P11_parameters}, since they are used 
in the RT calculations we present below. The scaling of the 
dust density double exponential distributions is determined by the face-on central optical depth in the B-band $\tau_B^{f}$. For the galaxy model of P11, a fixed ratio 
equal to 
0.387 is assumed for the face-on central optical depths of the thick dust disc and thin dust disc (see Eqn.\,10 in P11). The assumed 
opacity 
and scattering coefficients are those 
for $R_V$=3.1 from Weingartner \& Draine (2001), revised by Draine \& Li (2007). Note that, unlike the comparison with the \textsc{DUSTY} 
code, the scattering is considered anisotropic in this case. \\

For the comparison between the solutions provided by the two codes, we used the library 
of radiation fields presented in 
Popescu \& Tuffs (2013, PT13). From the library we obtained the contribution of the old stellar disc and young stellar disc to the 
radiation field 
energy density distribution separately, thus allowing a more careful check on the accuracy of the 3D calculation. \\

As for the comparison with the \textsc{DUSTY} code, the first step is to create the grids used in the calculations. We input in the grid creation
algorithm (see \S3.1) the functional shapes for the stellar and dust discs and impose the following 
criteria for the cell subdivision: 1) a maximum value for cell optical depth; 2) a maximum value for the cell luminosity; 3) a maximum 
cell subdivision level (typically about 6-7); 4) a subdivision level equal to the maximum subdivision level for the cells in the region close to 
the galaxy centre ($|z| < 1000$ and $R < 500$). By using condition 4, we required a higher spatial resolution in the galaxy central 
regions. Those regions are the ones where we expect 
the radiation field to vary more rapidly. An alternative way to obtain an increase of the spatial resolution in those regions would be by 
lowering further the threshold values for e.g. the cell optical depth required for the cell subdivision. However, this usually results 
in a substantial increase of the total number of cells even for relatively small changes for the required thresholds. For simplicity we 
decided to impose a condition simply based on the distance from the axis origin. 

For each leaf cell we assigned for both the dust density and stellar volume emissivity the values averaged over the cell volume. That is, we 
numerically integrated the double exponential functions $f(R,z)$ describing the stellar and dust distribution within the cell volume 
$V_c$ and then divided by $V_c$. That is:
\begin{equation}
 <f_c>=\frac{\int_{V_c}{f(R,z)dxdydz}}{V_c}
\end{equation}

As mentioned above, we performed 3D calculations for the models including either the old stellar disc or the young stellar disc. 
The details and results of these tests are described in the following two subsections.  

\subsubsection{Calculations for the old stellar disc}

We considered the old stellar disc of the P11 galaxy model for $old=1$ and we calculated the RFED 
distribution in the B-band for the following models (maximum and average cell optical depth and luminosity in Table \ref{table_P11_test_grid}. For all the 
models we used $f_U=10^{-7}$, $N_{\rm{rays}}=2$ and $f_L=10^{-3}$)\footnote{The choice of $f_U$ and $N_{\rm rays}$ is 
such to guarantee global energy balance within a few percent accuracy (see \S4).}:\\
\\
1) $\tau_B^f=0$ (optically thin model) \\
2) $\tau_B^f=1$, only the thick dust disc \\
3) $\tau_B^f=1$, both the thick and thin dust discs \\

\begin{figure}
\includegraphics[scale=0.7]{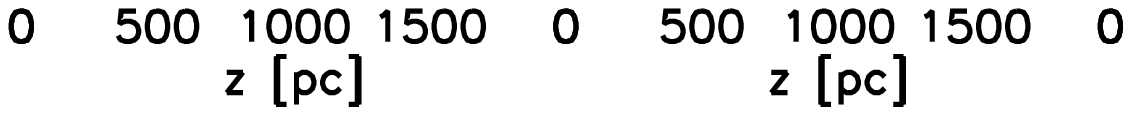}
\includegraphics[scale=0.7]{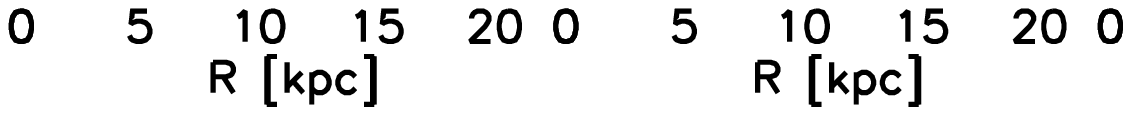}
\caption{RFED radial and vertical profiles for the old stellar disc in the optically thin case, B band. Diamonds represent the RFED values
calculated by our 3D code while the continuous line is the solution obtained by the 2D code of Popescu et al. (2011). The diamond symbols
plotted in the $R=0$ and $z=0$ plots represent the exact values obtained by the 3D code. For all the other plots, the plotted values are 
obtained through interpolation within the 3D grid.} 
\label{md_tau0_B}
\end{figure}

Model 1) has been calculated to check the accuracy of the 3D code when dust is not included. Fig.\ref{md_tau0_B} shows a comparison of 
the RFED radial and vertical profiles obtained by this calculation (diamonds symbols) and by the 2D code 
used by P11 (continuous line, same convention hereafter). For the comparison we selected the vertical profiles at $R=0, 5,10,18$\,kpc and 
radial profiles at $z=0,400,1000,2000$\,pc (note that while the plots for $R=0$ and $z=0$\,pc show the exact values obtained from the 3D 
calculation, the other plots show values obtained through interpolation within the 3D grid). As shown in Table \ref{sigma_popescu_11}, 
the average discrepancy for the 
vertical profiles is of order of 1-2\% for $R=0,5,10$\,kpc and about 5\% for $R=18$\,kpc. For the radial profiles the discrepancies are between
2-5\%, although most of the disagreement is found at large radial distances.\\

\begin{figure}
\includegraphics[scale=0.7]{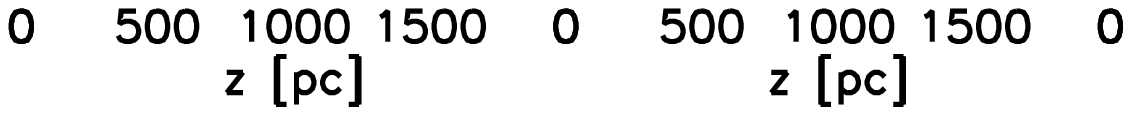}
\includegraphics[scale=0.7]{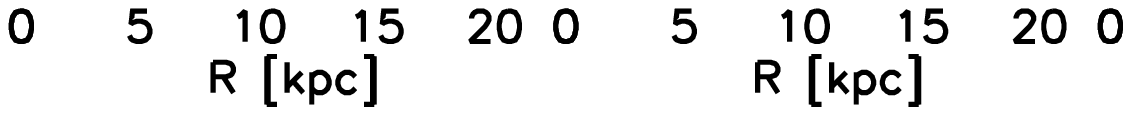}
\caption{RFED radial and vertical profile for the old stellar disc for $\tau_B=1$, B band. In this model the thin dust disc is not 
included. Also, only the contribution from direct stellar light is considered. Same symbols as in Fig.\ref{md_tau0_B}.} 
\label{md_tau1_B_notdust_dir}
\end{figure}

\begin{figure}
\includegraphics[scale=0.7]{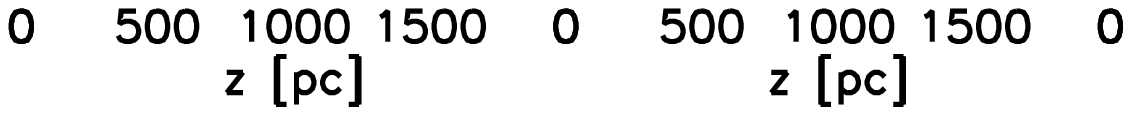}
\includegraphics[scale=0.7]{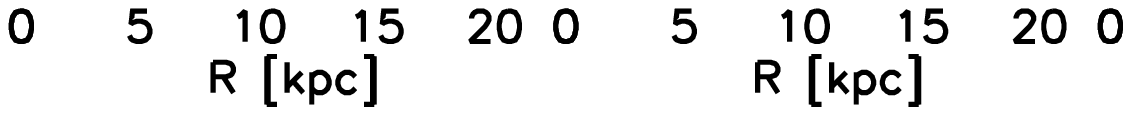}
\caption{RFED radial and vertical profile for the old stellar disc for $\tau_B=1$, B band. In this model the thin dust disc is not 
included. Also, only the contributions from direct stellar light and the first order scattered light are included. Same symbols as in Fig.\ref{md_tau0_B}.} 
\label{md_tau1_B_notdust_sca1}
\end{figure}

\begin{figure}
\includegraphics[scale=0.7]{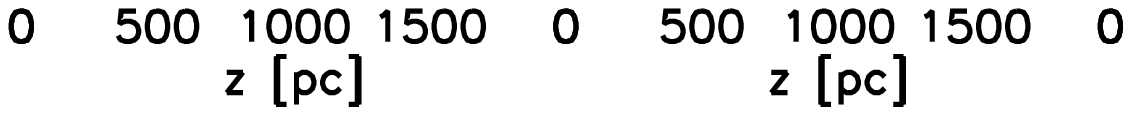}
\includegraphics[scale=0.7]{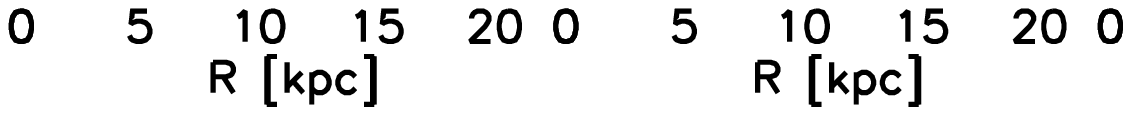}
\caption{RFED radial and vertical profile for the old stellar disc for $\tau_B=1$, B band. In this model the thin dust disc is not 
included. Both direct stellar light and all order scattered light contributions are included. Same symbols as in Fig.\ref{md_tau0_B}.} 
\label{md_tau1_B_notdust}
\end{figure}

Model 2) includes only the thick dust disc but not the thin dust disc. We calculated this model because both the old stellar disc and the
 thick dust disc can 
be well resolved by an adaptive grid containing of order of $10^5$ cells. Resolving the thin dust disc properly requires an order of 
magnitude more cells and this is avoided in this calculation to reduce the calculation time (but see model 3 below). 
Because the thin dust disc is not present, all the opacity is assigned to the thick dust disc (at variance with the original P11 galaxy model). 
For this test we show three sets of results. Fig.\ref{md_tau1_B_notdust_dir} shows the profile of the radiation field including only 
the direct light from the stellar distribution. Fig.\ref{md_tau1_B_notdust_sca1} includes the direct light and the first order scattered
light. This calculation has been performed since the 2D code of P11 is based on an algorithm which explicitly calculates 
only the first order scattered 
light, assuming that the ratio between the specific intensity of successive scattering orders is constant (see Kylafis \& Bahcall 1987 
for more details).  
Thus, it is important to verify the agreement at this stage before comparing the final solution including all order scattering, 
 shown in Fig.\ref{md_tau1_B_notdust}. The calculations including only direct light agree within 1-4\% for both vertical and radial 
profiles. The agreement for the
solution including also the first order scattered light is within about 0.5-2\% for the vertical profiles and within 1-3\% for the 
radial profiles. 
Finally, the comparison for the solution including all order scattered light shows an agreement within 1-2\% for the vertical profiles 
 and 2-3\% for the radial profiles. \\

\begin{figure}
\includegraphics[scale=0.7]{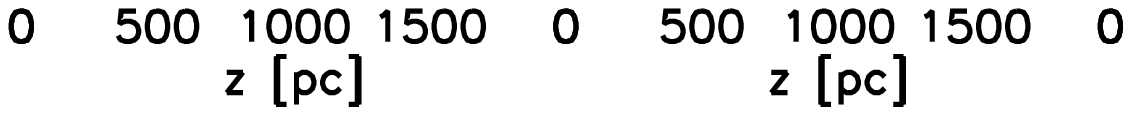}
\includegraphics[scale=0.7]{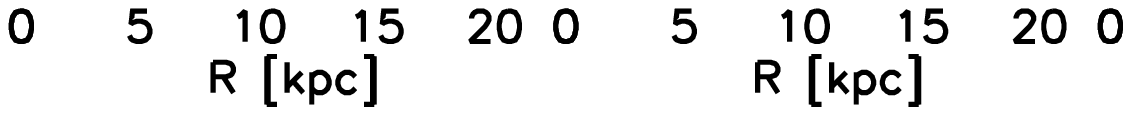}
\caption{RFED radial and vertical profile for the old stellar disc for $\tau_B=1$, B band with both the thick and thin disc included. 
Only direct stellar light is included. Same symbols as in Fig.\ref{md_tau0_B}.}
\label{md_tau1_B_dir}
\end{figure}

\begin{figure}
\includegraphics[scale=0.7]{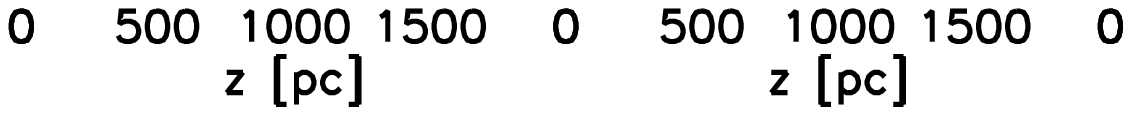}
\includegraphics[scale=0.7]{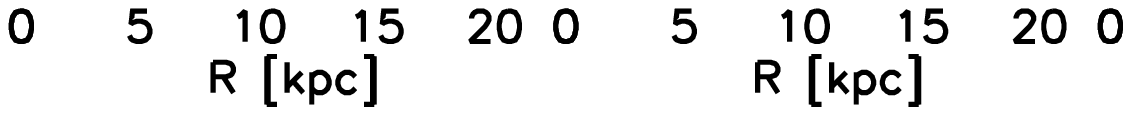}
\caption{RFED radial and vertical profile for the old stellar disc for $\tau_B=1$, B band with both the thick and thin disc included. 
Only direct stellar light and first 
order scattered light are included. Same symbols as in Fig.\ref{md_tau0_B}.}
\label{md_tau1_B_sca1}
\end{figure}

\begin{figure}
\includegraphics[scale=0.7]{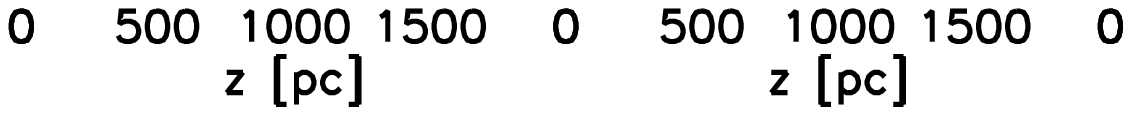}
\includegraphics[scale=0.7]{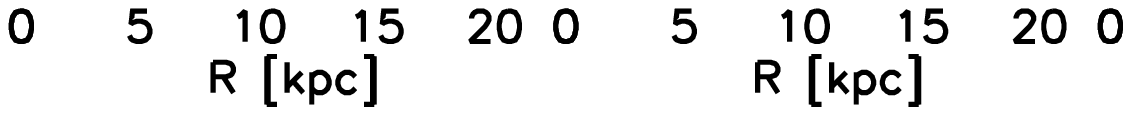}
\caption{RFED radial and vertical profile for the old stellar disc for $\tau_B=1$, B band with both the thick and thin disc included. 
Both direct stellar light and all order scattered light contributions are included. Same symbols as in Fig.\ref{md_tau0_B}.}
\label{md_tau1_B}
\end{figure}

For model 3) the calculations includes both dust discs, although the thin dust disc is not well resolved especially at large radii (cell
sizes on the galaxy plane varying between 65 and 592pc, while the scaleheight of the thin dust disc is about 91 pc). Similarly as before, 
Fig.~\ref{md_tau1_B_dir}, \ref{md_tau1_B_sca1} and
\ref{md_tau1_B} show the solutions for the RFED profiles including the direct light, the direct light plus the first order scattered light and 
the final values including all scattered light respectively. The comparison for the direct light shows an agreement within about 2-6\%
while the calculations including direct light and first order scattered light agree within 1-4\%.   
For the last calculation including all the stellar and scattered light contributions to the RFED, the average discrepancies are within 1-4\% for the vertical profiles,
about 7\% for the radial profile at $z=0$\,pc and about 3-4\% for the radial profiles at $z=400,1000, 2000$\,pc. \\

The 3D calculations performed for the old stellar disc show a good agreement with the 2D solutions, with the residual discrepancies being 
plausibly due to the resolution of the 3D grid which does not resolve properly the thin dust disc in the central regions and both the 
discs at large radii.

\begin{table}

 \begin{center}
 \begin{tabular}{lc}
\hline
\hline
$h_s^{\rm disc}$  &  5670 pc \\ 
$z_s^{\rm disc}$  &  419.58 pc \\ 
$h_s^{\rm tdisc}$ & 5670 pc \\ 
$z_s^{\rm tdisc}$ &  90.72 pc \\
$h_d^{\rm disc}$  & 7972.02 pc \\ 
$z_d^{\rm disc}$  & 272.16 pc \\
$h_d^{\rm tdisc}$ &  5670 pc \\ 
$z_d^{\rm tdisc}$ &  90.72 pc \\
$L_\nu^{\rm disc}$ &  4.771$\times$10$^{21}$\,W/Hz \\ 
$L_\nu^{\rm tdisc}$ & 2.271$\times$10$^{21}$\,W/Hz \\\hline
 \end{tabular}
\end{center}

\caption{Geometrical parameters of the disc and thin disc, together with the corresponding luminosity parameters 
for $old=1$ and $SFR=1 M_{\odot}/yr$. All the values for the stellar discs are those for the B-band. 
$h_s^{\rm disc}$ and $z_s^{\rm disc}$: scale height/length for the old stellar 
disc; $h_s^{\rm tdisc}$ and $z_s^{\rm tdisc}$: scale height/length for the young stellar disc;  $h_d^{\rm disc}$ and 
$z_d^{\rm disc}$: scale height/length for the thick dust disc; $h_d^{\rm tdisc}$ and $z_d^{\rm tdisc}$: scale height/length for the 
thin dust disc; $L_\nu^{\rm disc}$ and $L_\nu^{\rm tdisc}$: luminosity density for the old and young stellar disc respectively.} 
\label{table_P11_parameters}
\end{table}

\begin{table}

 \begin{center}
 \begin{tabular}{lccccc}
model & $\Delta\tau_{B,\rm{max}}$ & $<\Delta\tau_B>$ & $\Delta L_{B,\rm{max}}$ & $<\Delta L_B> $ & N$_{\rm cells}$\\ 
      &                      &                  &   $10^{17}$ W/Hz   &  $10^{17}$  W/Hz &  \\ \hline
 OLD (1,3) & 0.4             & 0.047            &   4.76             &   0.46          &  109620    \\
 OLD (2)  & 0.28             & 0.048            &   4.76             &   0.50          &  97200    \\
 YOUNG (1,2) & 0.4           & 0.059            &   2.27             &   0.25          &  95040   \\
 \end{tabular}
\end{center}

 \caption{3D grid parameters for the comparison with the 2D calculations of P11. Col.1: stellar population considered in the calculation, 
 OLD refers to the old stellar disc and YOUNG to the young stellar disc; col.2-3: maximum and average cell optical depth; col.4-5: 
 maximum and average cell luminosity; col.6: total number of cells.} 
\label{table_P11_test_grid}
\end{table}

\subsubsection{Calculations for the young stellar disc}

3D calculations to obtain the RFED distribution due to the thin stellar disc are more challenging because, as previously stated,  
in order to properly resolve the young stellar and thin dust discs, one should in principle create a grid containing of order of $10^6$ 
cells. 
We preferred to use a grid of about $10^5$ cells which underresolve the thin discs but it allows shorter computational times. 
The downside is that the 3D calculations will be less accurate numerically. \\  

\begin{figure}
\includegraphics[scale=0.7]{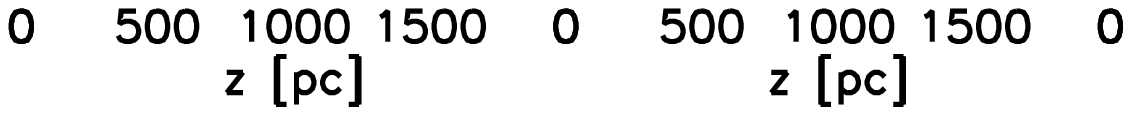}
\includegraphics[scale=0.7]{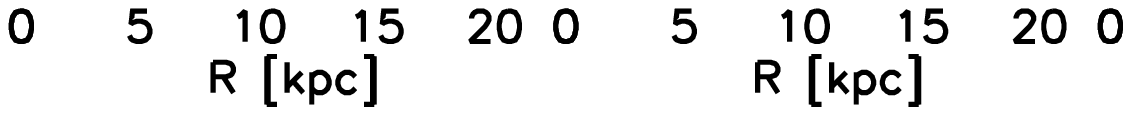}
\caption{Radial and vertical profile for the young stellar disc in the optically thin case, B band. Same symbols as in Fig.\ref{md_tau0_B}.} 
\label{mtd_tau0_B}
\end{figure}

For this set of tests we considered a young stellar disc with $SFR=1\,M_{\odot}/yr$ and we calculated the RFED in the B-band
for the following models (as before, we used $f_U=10^{-7}$, $N_{\rm{rays}}=2$ and $f_L=10^{-3}$): \\
\\
1) $\tau_B=0$ (optically thin case) \\
2) $\tau_B=1$, both dust discs\\
\\
Fig.\ref{mtd_tau0_B} shows the results for model 1). The calculated vertical profiles from the 2D and 3D codes agree within 1-4\% for 
$R=0,5,10$kpc while the average discrepancy is about 10\% for $R=18$kpc (see Table \ref{sigma_popescu_11}). The discrepancies tend to 
be higher for the radial profiles, that is, about 5-10\% with most of the discrepancy found at large radii. 
However, in this case the discrepancy seems not to be due only to the 
coarse 3D grid resolution at large radii. In fact, the 2D calculation show an artificial feature on the $R=18$\,kpc plot for low values 
of $z$. This feature is due to inaccuracies in the 2D calculation. \\

\begin{figure}
\includegraphics[scale=0.7]{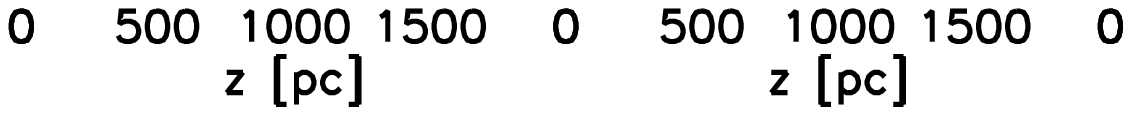}
\includegraphics[scale=0.7]{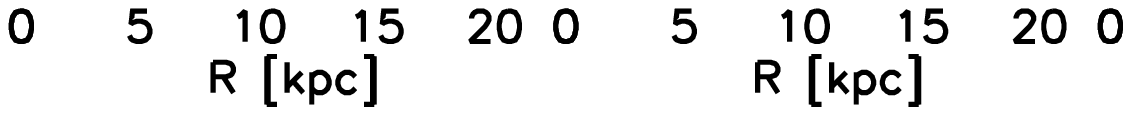}
\caption{Radial and vertical profile for the young stellar disc for $\tau_B=1$, B band. Same symbols as in Fig.\ref{md_tau0_B}.} 
\label{mtd_tau1_B}
\end{figure}

Fig.\ref{mtd_tau1_B} shows the calculations for model 2). The average discrepancies are of order of 3-6\% for the vertical profiles and 
about 5-7\% for the radial profiles. Note that the ``steps'' appearing in the radial profile 
for $Z=0$ are located at the position where the adaptive grid changes cell size. They appear also in the old stellar disc solutions but 
in a less evident way. \\

The 3D calculations for the young stellar disc provide solutions which still present a fairly good agreement, although there is
a higher discrepancy compared to the old 
stellar disc calculations. As said before, this is most probably mainly due to the low resolution of the 3D grid, unable to properly 
resolve the stellar and dust components in this test. 

\begin{table}
\begin{center}
\begin{tabular}{cccccccccc}
Model & Type & \multicolumn{8}{c}{$<\frac{\Delta U_\lambda}{U_\lambda}>$} \\
  & & $R_1$ & $R_2$ & $R_3$ & $R_4$ & $z_1$ & $z_2$ & $z_3$ & $z_4$\\\hline
OLD (1) & NODUST & 0.009 & 0.02 & 0.02 & 0.048 & 0.017 & 0.046 & 0.036 & 0.036   \\
OLD (2) & DIR & 0.014 & 0.018 & 0.019 & 0.037 & 0.038 & 0.03 & 0.016 & 0.028    \\
OLD (2) & DIR+SCA1 & 0.019 & 0.008 & 0.006 & 0.022 & 0.014 & 0.029 & 0.017 & 0.024 \\
OLD (2) & ALL & 0.016 & 0.015 & 0.015 & 0.013 & 0.02 & 0.022 & 0.034 & 0.032  \\
OLD (3) & DIR & 0.03 & 0.04 & 0.054 & 0.068 & 0.0079 & 0.06 & 0.034 & 0.027  \\
OLD (3) & DIR+SCA1 & 0.014 & 0.01 & 0.02 & 0.034 & 0.027 & 0.036 & 0.021 & 0.023  \\
OLD (3) & ALL & 0.036 & 0.013 & 0.01 & 0.025 & 0.073 & 0.027 & 0.035 & 0.035  \\ 
YOUNG (1) & NODUST & 0.0078 & 0.02 & 0.043 & 0.095 & 0.055 & 0.097 & 0.066 & 0.027   \\
YOUNG (2) & ALL & 0.052 & 0.061 & 0.046 & 0.032 & 0.069 & 0.065 & 0.054 & 0.048  \\

\end{tabular}
\end{center}
\caption{Average relative discrepancies for the B-band RFED profiles calculated using our 3D code and those from the code of P11. 
 Col. 1) model calculated (see text); col. 2) Type of calculation: NODUST=optically thin case, DIR=only direct stellar light, DIR+SCA1=only direct light 
 and first order scattered light, ALL=direct and all order scattered light; col.3-6: average discrepancies for the RFED vertical profiles
 at R$_1$=0\,kpc, R$_2$=5\,kpc, R$_3$=10\,kpc, R$_4$=18\,kpc; col. 7-10: average discrepancies for the RFED radial profiles at z$_1$=0\,kpc,
 z$_2$=400\,pc, z$_3$=1\,kpc,z$_4$=2\,kpc}
\label{sigma_popescu_11}
\end{table}

\clearpage

\section{An RT 3D application: including spiral arms in galaxy models}

For most practical applications to large statistical samples of galaxies (e.g. Driver et al. 2007, 2008, 2012, 
 Gunawardhana et al. 2011, Silva et al. 2011, Grootes et al. 2013), RT modelling of the observed spatially integrated 
 direct and dust-reradiated starlight necessarily (in the absence of detailed images) adopts 2D axisymmetric approximation of the 
 distribution of stars and dust. However, the distribution of light at UV and short optical wavelengths from young massive stars is 
 well known in
real galaxies to be biased towards a spiral pattern of enhanced dust density, rather than the smooth exponential disc function 
typically assumed by these models. The question therefore arises whether this effect introduces any systematic bias into 2D model 
predictions of dust attenuation of integrated starlight and averaged RFED in spiral galaxies. 
To evaluate this bias, we have performed a RT calculation for a galaxy 
model including logarithmic spiral arms. We considered a typical model 
galaxy from P11, consisting of a disc with $old=1$ and a thin stellar disc with $SFR=1M_\odot/yr$, and two dust
discs with $\tau^f_B=1$ (see Sect. 5.2 for a description of the model parameters of P11). We modified this model using the same 
procedure as that adopted in P11 for the inclusion of circular spiral arms.  Thus, we considered 
the same double exponential distribution for the thick stellar and dust disc but we redistributed the thin disc stellar luminosity and 
 dust mass within spiral arms. As shown in Schechtman-Rook et al. (2012), the implementation of logarithmic spiral arms can be done 
 by multiplying a logarithmic spiral disc perturbation $\xi$ to the double exponential formula describing the stellar volume emissivity 
 and dust density (see Eqn. \ref{double_exp}). We adopted the expression for $\xi$ in their formula 10 (two spiral arms): 
\begin{equation}
 \xi=\left[1-w + \prod_{n=2,n+2}^N \frac{n}{n-1} \sin^N\left(\frac{\ln(\sqrt{x^2+y^2})}{\tan(p)}-\tan^{-1}(\frac{y}{x})+\frac{\pi}{4} \right) \right]
\end{equation}
with $w$ the fraction of stellar light or dust within the spiral arms, $p$ the pitch angle determining how tightly the spirals turn
around each other and the exponent $N$ which regulates the relative size of the arm and interarm regions. For these parameters, we adopted
the values $w=0.9$, $p=10\,^{\circ}$ and $N=10$. Then, we performed an RT calculation for a galaxy model (including both old and young 
stellar discs) in the B-band with face-on central optical depth $\tau_B^f=1$ and disc parameter values as in Table 
\ref{table_P11_parameters}.

\begin{figure}
\centering
\includegraphics[scale=0.3]{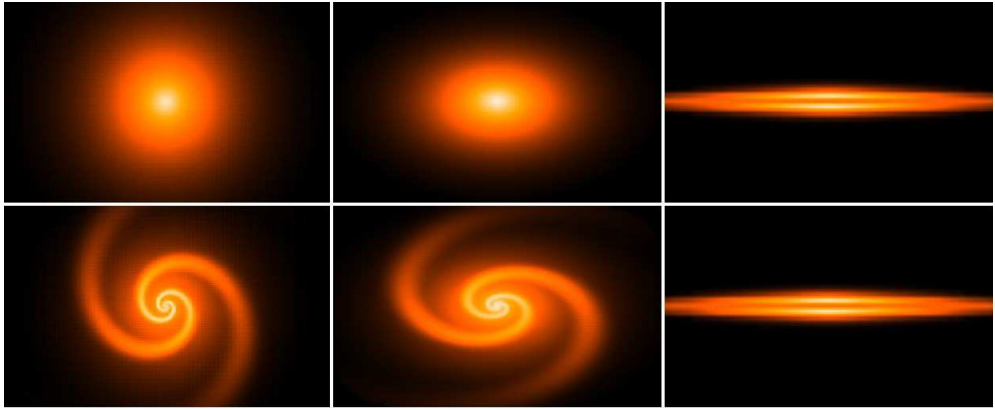}
\caption{Comparison of the images of output surface brightness at different inclinations ($0^\circ$, $51^\circ$ and $90^\circ$ 
from left to right) for a pure double exponential disc galaxy model (upper row) and a model including spiral arms (lower row). The models are for a central face-on optical depth $\tau_B^f=1$. 
See text for details.} 
\label{spiral_model_comparison}
\end{figure}

\begin{figure}
\centering
\includegraphics[scale=0.4]{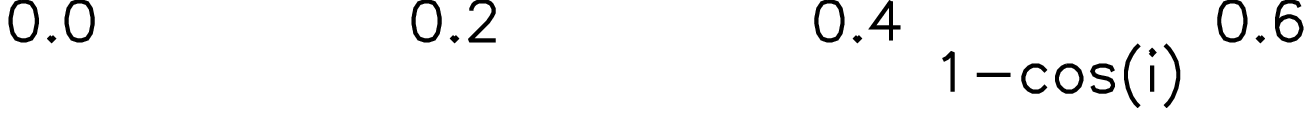}
\caption{Comparison of the attenuation curves in the B-band as a function of galaxy inclination for the pure double exponential model
(continuous line) and for the model with spiral arms (dashed line)} 
\label{plot_att_model}
\end{figure}

Fig.\ref{spiral_model_comparison} shows the comparison for the output surface brightness images at different inclinations 
between a pure double exponential model (upper row) and for the model including spirals (lower low). The images show the different 
morphology of the stellar emission for the face-on and low-inclination images. However, the edge-on images are remarkably similar for 
the two models. We also made a comparison for the total attenuation as a function of galaxy inclination for the two models, which is shown in 
 Fig.\ref{plot_att_model}. The attenuation curves are quite close to each other, within 0.02 dex, showing that the spiral pattern 
 does not affect much the total attenuation of the galaxy for the adopted parameters. 
 
\begin{figure}
\centering
\includegraphics[scale=0.4]{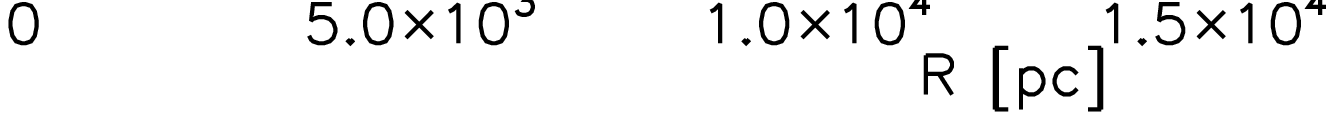}
\caption{Comparison of the B-band RFED radial profile for the models with and without spiral arms. The squares 
represent the RFED values for the pure double exponential galaxy model. The blue line is the RFED profile along the 
x-axis of the spiral galaxy model, while the red line is the RFED profile azimuthally averaged for the same model. } 
\label{plot_urad_prof_model}
\end{figure}

Finally, we compared the RFED profiles in the galaxy plane. Fig.\ref{plot_urad_prof_model} shows the profiles for the pure double exponential 
model (squares), a cut along the x-axis of the model including spiral arms (blue line) and its azimuthally averaged RFED profile.
Interestingly enough, although the RFED along the x-axis shows the variation due to the spiral arms, the azimuthally averaged profile 
is very close to the profile for the model without spiral arms.

Although P11 have already shown that the spatially integrated dust and PAH emission SED of a typical spiral galaxy 
does not depend on whether the young stellar population and associated dust is distributed in a circular spiral arm or in a disc, 
here we show for the first time that it is at the level of the radiation fields that heat the dust that the distributions in the 
spirals start to resemble the disc distributions on the average.

The results on the global attenuation, images and radiation fields, all suggest that double exponential models can be a quite good 
representation for spiral disc galaxies, and that the spatially integrated SEDs of spirals can be accounted by 2D models. 
Although this is in qualitative agreement with previous works (Misiriotis et al. 2000, Semionov et al. 2006, Popescu et al. 2011), a more extensive 
study is needed to see how the different parameters, e.g. the face-on optical depth, affect the attenuation curve and the 
radiation fields in models with and without spiral arms.  

\clearpage

\section{Summary and Outlook}

In this paper we present a new ray-tracing dust radiation transfer algorithm which is able to handle arbitrary 3D geometries and 
it is specifically designed to calculate accurate RFED within galaxy models. The main optimization characteristics of this algorithm are 
the following:\\
 1) an adaptive 3D grid (see \S3.1)\\
 2) a ray-tracing algorithm based on the pre-calculation of a lower limit for the RFED (see \S3.2 and \S3.3)\\ 
 3) an iterative procedure for the optimization of the angular density for the rays departing from each emitting cell (\S3.3).\\
Furthermore, parallelized versions of the code have been written for its use on shared-memory machines and computer clusters.\\ 

In order to verify the code accuracy, we performed comparisons with the results provided by other codes. 
Specifically, we used our code to calculate solutions for a spherical dusty 
shell illuminated by a central point 
source and for an axis symmetric galaxy model. For the first configuration, we considered as 
benchmark a set of solutions calculated by using the \textsc{DUSTY} 1D code (Ivezic \& Elitzur 1997). We showed that the equilibrium dust  
temperature radial profiles and the outgoing flux spectra derived by our 3D calculations agree within a few percent with the benchmark 
solutions for models with low radial 
optical depths ($\tau_1=1,2$) but present larger discrepancies for more optically thick models ($\tau_1=5,10$). The residual 
discrepancies, especially for the models with higher optical depths, are most probably due to the lack of dust self-heating in our code 
and the lower spatial resolution of the 3D calculations compared to the 1D ones. Since the geometry of the source emission/opacity of star/dust shell
does not reproduce that for which our algorithm was developed, namely that of an extended distribution of stellar emission and dust, 
we also used a second benchmark. Thus, we considered the 2D calculations 
by Popescu et al. (2011) for the RFED distribution within their galaxy model. We calculated
the contribution to the RFED distribution due to an old stellar disc and a young stellar disc separately and we compared the results 
for radial and vertical RFED profiles derived for a set of reference radii and vertical distances.
We found a general good agreement between the 3D and 2D calculations within a few percent in most of the cases. At least part of the 
residual discrepancy can be accounted by the relatively low spatial resolution of the grid used in the 3D calculation, 
which is not sufficient to properly resolve the thin disc component of the galaxy model of Popescu et al. (2011).
We showed an example of a 3D application of the code by performing RT for a spiral galaxy model where in one case the 
emissivity of the young stellar population and associated dust opacity are distributed in logarithmic spiral arms and in another case 
are distributed in exponential discs. We found that the edge-on images, the attenuation as a function of inclination and 
the azimuthally average RFED profiles on the galaxy plane are approximately the same for the two models.
This suggests that the spatially integrated SEDs of spirals can be well described by 2D models.

The tests we performed have shown that, in the conditions where dust self-heating is negligible and 
the 3D spatial resolution is high enough to resolve emission and opacity distributions, our code can be used to calculate 
 accurate solutions for the RFED. This characteristic is particularly important for the calculation of stochastically heated dust 
 emission, which requires both the overall intensity and the colour of the radiation field to be calculated in an accurate way. 
In a future work we will show applications of the code for the calculation of infrared emission. This will be performed using our 3D 
RT code coupled with the dust emission code used by Popescu et al. (2011), which 
self-consistently calculates the stochastic emission from small grains and PAH molecules. In this way, it will be possible to use our 
code to obtain both integrated 
SEDs and images in the mid- and fir-infrared for galaxies with arbitrary geometries.  
In addition, an important step will be to further optimize the code in order to make it possible to run on grids containing millions of 
cells in a reasonably short time. This will allow us to 
improve further the accuracy of the calculation for RFED within multi-scale structures spanning at least three orders of 
magnitude, such as from $\approx10$\,pc to $\approx$10\,kpc in the case of a galaxy ISM. \\

\clearpage

\section*{Acknowledgements}
We acknowledge support from the UK Science and Technology Facilities Council (STFC; grant ST/J001341/1). GN thanks S. Dalla,
K. Foyle, E. Kafexhiu, T. Laitinen, J. Steinacker for useful suggestions and/or discussions. CCP thanks the Max 
Planck Institute f\"{u}r Kernphysik for support during a sabbatical, when this work was completed.

\appendix

\end{document}